\documentclass[aps,twocolumn,amsfonts,amsmath,amssymb, preprintnumbers]{revtex4} 

\usepackage{graphicx}
\usepackage{dcolumn}  
\usepackage{bm}  

\begin{document}
\setlength\arraycolsep{2pt}

\title{Diffusion on Ruffled Membrane Surfaces}

\author{Ali Naji \email{E-mail: anaji@chem.ucsb.edu}}
					
\affiliation{Department of Physics, and Department of Chemistry and Biochemistry,
University of California, Santa Barbara, CA 93106}
                                
\author{Frank L. H. Brown \email{E-mail: flbrown@chem.ucsb.edu}}

\affiliation{Department of Chemistry and Biochemistry, University of California, Santa Barbara, CA 93106}

\preprint{Published in J. Chem. Phys. {\bf 126}, 235103 (2007)}

\begin{abstract}
We present a position Langevin equation for overdamped particle motion on
rough two-dimensional surfaces.  A Brownian Dynamics algorithm is suggested to evolve
this equation numerically, allowing
for the prediction of effective (projected) diffusion coefficients over corrugated surfaces.
In the case of static surface roughness, we find that a simple area-scaling prediction for the projected
diffusion coefficient leads to seemingly quantitative agreement with numerical results.
To study the effect of dynamic surface evolution on the diffusive process, we consider particle diffusion over
a thermally fluctuating elastic membrane.  Surface fluctuation has the effect of increasing the effective 
diffusivity toward a limiting annealed-surface value discussed previously.
We argue that protein motion over cell surfaces spans a variety of physical regimes, 
making it impossible to identify a single approximation scheme appropriate to all measurements of interest.
\end{abstract}

\maketitle



\section{Introduction}
\label{sec:intro}

Diffusive processes are commonplace in many systems of chemical, physical and
biological interest \cite{crank,berg,Risken,vankampen}, a fact commonly attributed to inherent randomness at
the molecular level of dynamics (thermal or otherwise) and the wide range of applicability of the central limit
theorem \cite{vankampen}. In the context of biological systems, diffusion plays a key role in many
of the processes of life at the cellular and sub-cellular levels \cite{berg,The-Cell}.  
In particular, the diffusion
of molecules on the surface of cells has received widespread attention for many years
\cite{purcell,singer,lipowsky,Jacobson95,Saxton97}, both due to the clear consequences this motion 
has for cellular functioning as well as the availability of suitable experimental techniques to make quantitative
measurements.  Developments in experimental techniques such as single-particle
tracking \cite{Saxton97},  
fluorescence photo-bleaching recovery \cite{Axelord}
and nuclear magnetic resonance (NMR) \cite{Halle_NMR} have made
it possible to gain direct insight into the structure and dynamics of lipid membranes 
by investigating the lateral diffusion of lipids and integral membrane proteins \cite{Jacobson93}. 

\begin{figure}[t!]
\begin{center}
\includegraphics[angle=0,width=8cm]{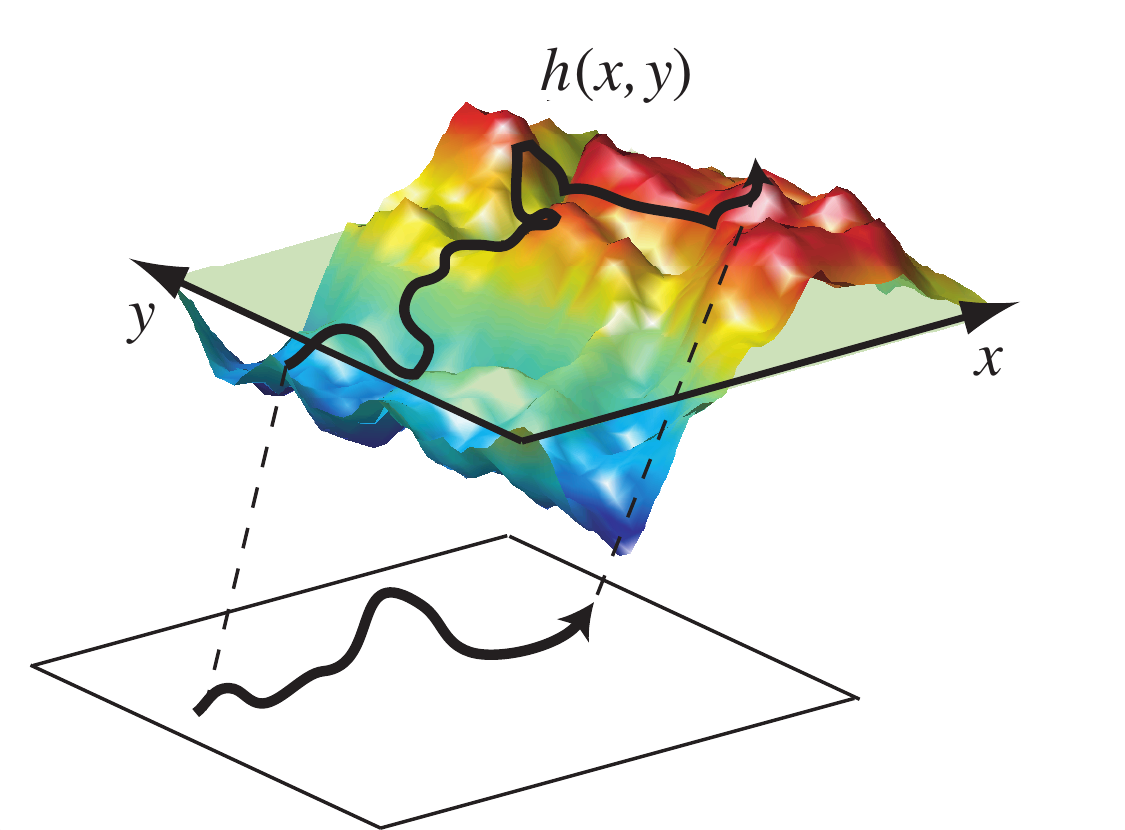}
\caption{A typical equilibrium configuration of a semiflexible elastic membrane described in Monge gauge. 
Projected self-diffusion in the $x-y$ plane (with effective diffusion coefficient $D$) is necessarily 
reduced relative to the curvilinear self-diffusion over the actual surface (with diffusion coefficient $D_0$) 
since any path followed on the membrane surface projects to a path of equal or smaller length in the base plane.}
\label{fig:path_projection}
\end{center}
\end{figure}

Biological membranes are typically not flat, which
leads to ambiguity in assigning the ``diffusion coefficient'' for a molecule on such a surface.  
Do we mean the intrinsic curvilinear diffusion coefficient, $D_0$, that would be expected if 
the molecule were moving over a truly flat surface, or do we mean the effective diffusion 
coefficient, $D$, reflecting projected motion of the particle onto a reference plane (see Fig. \ref{fig:path_projection})?  
Most experiments, including optical techniques and NMR \cite{Saxton97,Axelord,Halle_NMR}, 
directly measure $D$, while traditional theories provide a route to estimation of 
$D_0$ \cite{Saffmann,Broday}. The connection between $D$ and $D_0$ is the focus of this paper.

It is well appreciated that confining a particle to a two-dimensional (2D) surface within a 
three-dimensional (3D) space has the effect of lowering the effective diffusion coefficient, $D$, relative
to the curvilinear value, $D_0$ \cite{Halle, Halle_1d,Holyst,Faraudo,Christensen,Jackson,Festa}.  Recently, this
geometric effect has seen renewed interest in the context of diffusion over lipid bilayers and 
biomembranes \cite{King,Seifert,Gov,Sbalzarini}.  Reflecting the diversity of membrane 
structures present in biology, some studies have focused interest on the case of a particle diffusing 
over a static membrane \cite{Halle,King,Sbalzarini}, while others have concentrated on 
membranes with fast dynamic fluctuations \cite{Seifert,Gov}.  For certain biological structures, the static 
membrane (quenched disorder) limit is probably a good one.  For example, the endoplasmic 
reticulum (ER) \cite{The-Cell} and the lateral cortex of auditory outer hair cells \cite{Boal_Book,OHCs} are
oddly shaped in a predominantly static fashion due to topology of the membrane and interactions 
with other cellular components.  Other systems, like reconstituted unilamellar vesicles \cite{milner} and the 
plasma membrane of the red blood cell \cite{Brochard} are expected to fluctuate rapidly, which motivated 
Reister and Seifert to consider the fast shape-fluctuation (annealed disorder) limit in the context of membrane 
protein diffusion \cite{Seifert}.  The annealed disorder limit is especially attractive for theoretical study, since 
spatial correlations of the membrane surface drop out in this limit and analytical results are readily obtained \cite{Halle}.  
More generally, membrane systems may fall in an intermediate regime where membrane motion and diffusive transport 
share comparable characteristic time scales. 

While the annealed disorder limit is well understood, the opposite limit of quenched disorder is
less established.  One of the studies mentioned above \cite{Sbalzarini} introduces an advanced 
computational strategy to analyze real experimental data for static configurations of ER membranes, 
but avoids detailed theoretical interpretation of the results.  The static surface study by King \cite{King}, 
is plagued by an imprecise 
treatment of the diffusive process, which unfortunately leads to spurious results.  Gustafsson and Halle \cite{Halle} 
suggested a number of plausible approximations and rigorous bounds for the behavior of $D$ over a static surface, 
and hypothesized that an effective medium approximation (EMA) \cite{Solokov} would provide the closest 
correspondence with reality, but provided no direct evidence to support this conclusion.

Outside the field of biological physics, general simulation strategies have been suggested
for the evolution of diffusive particles over curved surfaces of arbitrary complexity \cite{Holyst,Christensen}.  
While these methods can be applied to membrane systems, in many cases of interest they are
more elaborate and computationally expensive than necessary.  Commonly, biomembrane surfaces are described 
in the Monge gauge.  The shape of the membrane surface is described by a single-valued function, 
$h(x,y)$, of lateral coordinates,  $x$ and $y$  \cite{single-value}; $h(x,y)$ provides the local height of the membrane 
above some arbitrary reference plane (see Fig. \ref{fig:path_projection}).  Given this parameterization, 
it would seem natural to evolve the position of a particle in terms of its $x$ and $y$ coordinates, 
without direct consideration of the third dimension (excepting the implicit information found in $h(x,y)$).

This paper presents a 2D Langevin equation for diffusive particle motion over a curved Monge-gauge 
surface that may be static or fluctuating in time.  
The numerical implementation of this equation leads to simulations of increased efficiency relative 
to the more general 3D random-walk methodology of Holyst et al. \cite{Holyst}.  
Simulations are carried out here for surfaces with lateral periodicity.

In the quenched limit, simulation results for a variety of differently-shaped surfaces (periodic in
a box of linear dimension $L$) suggest that 
the ratio between 
the projected and the curvilinear diffusion coefficients is equal to the ratio of projected membrane 
area, ${\mathcal A}_\bot=L^2$, to actual curvilinear membrane area, ${\mathcal A}$ (to within the precision we are able 
to achieve numerically).  That is,  $D/D_0 \sim {\mathcal A}_\bot/{\mathcal A}$ for static Monge-gauge surfaces.  
These results directly refute an earlier claim of non-integer scaling with surface area \cite{King} and disprove 
the suggestion \cite{Halle} that the effective medium approximation  should be the most accurate approximation available for predicting $D$ from $D_0$. 

To examine the effects of dynamic surface fluctuations, we consider Helfrich elastic membranes with
thermal dynamics described by a generalized continuum Langevin equation \cite{lipowsky}.  
As a general trend, we show that the ratio $D/D_0$ grows with increasing membrane fluctuation 
rates.  A particle moving on a slowly fluctuating membrane exhibits $D$ approaching the quenched
disorder limit discussed above, while fast membrane fluctuations effect  $D$ values
limiting to the somewhat larger annealed limit results discussed previously \cite{Halle,Seifert}.
For a membrane of given rigidity, the crossover between these two regimes is controlled by a dimensionless
parameter  $\chi = k_{\mathrm{B}}T/(4\eta L D_0)$, which represents  the ratio between
characteristic particle diffusion time and membrane shape-relaxation time scale in a hydrodynamic medium of
viscosity $\eta$.  A formal operator expansion in powers of $\chi^{-1}$ demonstrates the
pre-averaging Fokker-Planck equation \cite{Seifert} as the limiting behavior associated 
with fast membrane fluctuations ($\chi^{-1}\rightarrow 0$).  Additionally, this expansion 
suggests higher-order corrections, which may be of use in further theoretical study.  

We consider a wide range of $\chi$ values in our simulations to fully characterize the transition
between quenched and annealed regimes.  Typical membrane proteins in 
sufficiently large tensionless biomembranes are expected to exhibit intermediate $\chi$ values, with $D$ values 
falling in-between the quenched and annealed limits.



\section{Langevin equation for diffusion on curved surfaces}
\label{sec:langevin}

\subsection{Membrane geometry}
\label{subsec:geometry}

Let us briefly review some elementary geometric concepts of surfaces 
in order to introduce notation \cite{diff_geometry}.  
A curved 2D surface is parameterized by two independent coordinates.
In the Monge gauge, one makes use of  two Cartesian coordinates $(x, y)$ with respect to a 
fixed frame in a 3D Euclidean space; in the absence of overhangs \cite{single-value}, $z=h(x,y)$
uniquely represents the membrane height from the $x-y$ plane (see Fig.  \ref{fig:path_projection}). 
It is also convenient to define $\phi({\mathbf r})=z-h(x, y)$, to yield the surface equation 
in 3D: $\phi({\mathbf r})=0$ where ${\mathbf r}=(x, y, z)$. The gradient vector,  
$\nabla_{\mathbf r} \phi=(-\partial_x h, -\partial_y h, 1)$, points normal to the surface with magnitude
equal to the determinant of the surface metric 
\begin{equation}
  g= |\nabla_{\mathbf r} \phi|^2 = 1 + (\partial_x h)^2 + (\partial_y h)^2,  
\end{equation}  
which formally determines the area rescaling of an infinitesimal surface element upon projection
onto the $x-y$ plane, i.e.
\begin{equation}
{\mathrm{d}}\mathcal{A}=\sqrt{g}\,{\mathrm{d}}\mathcal{A}_{\bot}=\sqrt{g}\,{\mathrm{d}}x\,{\mathrm{d}}y, 
\label{eq:diff_area}
\end{equation}
where $\mathcal{A}$ represents the curvilinear area of the surface and $\mathcal{A}_{\bot}$,  the area projected 
on the $x-y$ plane.

Using a condensed notation $x_i=x, y$ for $i=1,2$, the
metric tensor elements of the surface in the Monge gauge read 
\begin{equation}
 g_{ij} = \delta_{ij} + \partial_i h \,\partial_j h, 
\end{equation} 
with $\delta_{ij}$ being the Kronecker delta and $\partial_i\equiv  \partial/\partial x_i$ \cite{Note_co_contra}. 
Hence one obtains the determinant $g={\mathrm{det}}(\mathbf{g})$ and the inverse 
tensor 
\begin{equation}
 (\mathbf{g}^{-1})_{ij} = \delta_{ij} - \frac{\partial_i h \,\partial_j h}{g}.  
\label{eq:g_inverse_monge}
\end{equation} 

Another useful quantity is the curvature tensor defined as \cite{diff_geometry}
\begin{equation}
 K_{ij} = \frac{\partial_i \partial_j h}{\sqrt{g}} = - \frac{\partial_i \partial_j \phi}{\sqrt{g}}.  
\end{equation} 
We will express some of our results in terms of the mean curvature,  
\begin{equation}
H = \frac{1}{2} (\mathbf{g}^{-1})_{ij}\, K_{ij}
\label{eq:mean_curvature}
\end{equation}
which gives the average of the two inverse principal radii of curvature at a given point on the 
surface. 
The Einstein summation convention is assumed above and in all that follows.

\subsection{Discrete random walk on a membrane}
\label{subsec:deriv_rw}

\begin{figure}[t!]
\begin{center}
\includegraphics[angle=0,width=7.5cm]{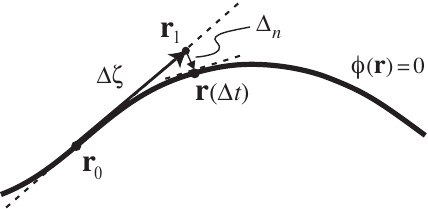}
\caption{Schematic representation of Eq. (\ref{eq:RW_eq}). Every discrete  random-walk step (of duration $\Delta t$)
over a curved surface (shown here by a thick curve for simplicity) may be broken down into two consecutive 
steps \cite{Holyst}. First, a random jump of size $\Delta {\boldsymbol \zeta}$ and duration $\Delta t$
from the initial position ${\mathbf r}_0$ to a new position  ${\mathbf r}_1$ in the local tangent plane (shown by dashed line).
Second, an instantaneous normal-projection move of size $\Delta_n$ from ${\mathbf r}_1$ to the final location, ${\mathbf r}(\Delta t)$,
on the surface. In the limit $\Delta t\rightarrow 0$, particle motion becomes exactly confined to the surface and, as shown
in the text, may be describe by a well-defined Langevin equation.  For the sake of representation, the jump size in the figure
is exaggerated; note that $\Delta_n$ is 
perpendicular to both dashed tangent lines
within the first order in $\Delta {\boldsymbol \zeta}$. }
\label{fig:RWschematic}
\end{center}
\end{figure}

Our goal is to establish a Langevin equation that describes the Brownian motion of a
particle confined to $z=h(x, y)$.  To this end, we first consider the case of discrete (but small)
displacements over the membrane surface.
Let us assume that a particle starts at ${\mathbf r}_0$ on the membrane surface (i.e.,
$\phi({\mathbf r}_0)=0$) before the discrete jump.  A Brownian jump of finite duration
$\Delta t$ may be broken down into two consecutive steps \cite{Holyst} (see Fig. \ref{fig:RWschematic}): 
i) The particle makes a random jump, $\Delta {\boldsymbol \zeta}=(\{\Delta \zeta_i\}, \Delta \zeta_z)$, 
to a new position  ${\mathbf r}_1={\mathbf r}_0+\Delta {\boldsymbol \zeta}$; this displacement is 
restricted to lie in the {\em local tangent plane} to the surface at ${\mathbf r}_0$ and is distributed according to 
a Gaussian probability distribution to be specified below. 
ii) The final location, ${\mathbf r}(\Delta t)$, 
of the particle is then obtained by an instantaneous normal projection of  ${\mathbf r}_1$ back
onto the membrane surface. 

The net displacement of the particle, 
$\Delta {\mathbf r} \equiv {\mathbf r}(\Delta t)-{\mathbf r}_0$,  
may thus be written as 
\begin{equation}
  \Delta {\mathbf r} = \Delta {\boldsymbol \zeta} +  \Delta_n
\label{eq:RW_eq}
\end{equation}
where the second term on the right hand side represents the normal projection (step ii), i.e.
\begin{equation}
    \Delta_n \equiv 
		{\mathbf r}(\Delta t)-{\mathbf r}_1 \simeq
			-\frac{\phi({\mathbf r}_1) \, \nabla_{{\mathbf r}_1} 
				\phi({\mathbf r}_1)}{|\nabla_{{\mathbf r}_1} \phi({\mathbf r}_1)|^2}, 
				\label{eq:projection}
\end{equation}
and for the first term we have
\begin{equation}
   \Delta {\boldsymbol \zeta}\cdot \nabla_{{\mathbf r}_0} \phi({\mathbf r}_0) = 0.
\label{eq:zeta_tangent}
\end{equation}
The above equations have been used to numerically simulate diffusion over a curved surface
\cite{Holyst}.  A significant merit of this methodology is that it may be used on arbitrarily shaped
surfaces; i.e., a single-valued Monge-gauge parameterization is not required to implement this random-walk algorithm.  
In the case of a surface that may be specified by $h(x,y)$, this procedure
may be simplified considerably, reducing the inherently 3D algorithm to a strictly 2D
equation of motion, with the surface constraint $z=h(x, y)$ met exactly. 
(In contrast, the surface constraint is satisfied 
only to second order in jump size within the random-walk algorithm as implied 
by Eq. (\ref{eq:projection}) \cite{Holyst}.)  This 2D equation of motion is specified in the following section.  


Let us consider the second term in Eq. (\ref{eq:RW_eq}). For sufficiently small $\Delta t$, 
the length of the initial jump, $\Delta {\boldsymbol \zeta}$, becomes small enough to justify
expanding  $\phi({\mathbf r}_1)$ around ${\mathbf r}_0$.  To second order in $\Delta {\boldsymbol \zeta}$, 
\begin{equation}
   \Delta_n  \simeq
				-\frac{1}{2} \left [ (\Delta{\boldsymbol \zeta}\cdot\nabla_{{\mathbf r}_0} )^2 \phi({\mathbf r}_0)\right ] \, 
					\frac{\nabla_{{\mathbf r}_0} \phi({\mathbf r}_0)}
						{|\nabla_{{\mathbf r}_0} \phi({\mathbf r}_0)|^2} 
							\,+\,{\mathcal O}\left(\Delta{\boldsymbol \zeta}^3\right), 
\end{equation}                                     
which shows that the projection displacement, $\Delta_n $, is of the second order in 
the random jump size, $\Delta {\boldsymbol \zeta}$.

Using the notation introduced in Section \ref{subsec:geometry}, and rewriting this equation in Monge gauge, 
one finds the following set of equations
\begin{equation}
  \Delta x_i = -\frac{1}{2} \bigg(\Delta \zeta_j \, \Delta \zeta_k \, \partial_j \partial_k h \bigg )
 				\frac{\partial_i h}{g}\bigg|_0 + \Delta \zeta_i 
					+{\mathcal O}\left(\Delta{\boldsymbol \zeta}^3\right) 
		\label{eq:dis_x_i}
\end{equation}
\begin{equation}	
  \Delta z = \frac{1}{2}  \bigg(\Delta \zeta_j \, \Delta \zeta_k \, \partial_j \partial_k h \bigg ) \frac{1}{g} \bigg|_0
					+ \Delta \zeta_z +{\mathcal O}\left(\Delta{\boldsymbol \zeta}^3\right). 
		\label{eq:dis_z}
\end{equation}
Recall that $i, j, k=1,2$ denote only the projected $x-y$ components. 

Note that in the first term on the right hand side of both equations,  only the $x-y$ components of 
the random jump $\Delta {\boldsymbol \zeta}$ enter. 
By virtue of Eq. (\ref{eq:zeta_tangent}), it follows that
the $z$-component of the random jump, $\Delta \zeta_z = \Delta \zeta_i \, \partial_i h$, is not an independent variable. 
As a result, for sufficiently small time steps or random jumps, 
the displacement in the $z$ direction, $\Delta z$, is fully described 
by the projected displacement in $x-y$ plane.  In other words, Eq. (\ref{eq:dis_z})
does not constitute an independent equation but merely a relation for the particle height 
variation, $\Delta z$, after each discrete step as enforced by the surface constraint $z=h(x, y)$. 
Indeed, one can easily check using Eqs. (\ref{eq:dis_x_i}) and (\ref{eq:zeta_tangent}), that 
 $\Delta z$ is simply equal to the variation of the membrane height, $\Delta h$, 
for a small projected jump $\Delta x_i$, i.e.
\begin{equation}
  \Delta z = \Delta h = \Delta x_i\, \partial_i h\big|_0 + \frac{1}{2} \Delta x_i \,\Delta x_j \,\partial_i \partial_j h\big|_0 +\ldots, 
\end{equation}
where the right hand side is a Taylor expansion. 
Hence,  in the limit of infinitesimal jumps, the particle is automatically confined to the surface as its $x-y$ position
evolves according to Eq. (\ref{eq:dis_x_i}).

\subsection{Continuous time limit: Langevin equation}
\label{subsec:deriv_langevin}

We now consider the time evolution relation (\ref{eq:dis_x_i}) in the limit of infinitesimal 
time steps, that is formally $\Delta t\rightarrow {\mathrm{d}}t$ 
or for the random jumps, $\Delta \zeta_i \rightarrow {\mathrm{d}}\zeta_i$. 
The random jump was chosen so that it locally describes a free Brownian motion 
 in the local tangent plane (characterized by a Gaussian distribution function) 
 with a curvilinear diffusion coefficient $D_0$ and duration $\Delta t$. Thus, 
the projected jump probability density may be written as
\begin{eqnarray}
    P\left(\{{\Delta \zeta}_i\}\right)\! &= &\frac{ \int\! {\mathrm{d}}{\mathbf r}\, G_0({\mathbf r})\,
    							\delta(z - x_i \partial_i h) \,\Pi_i \delta(\Delta \zeta_i -x_i) }
						{\int\! {\mathrm{d}}{\mathbf r} \,G_0({\mathbf r}) \,\delta(z - x_i\partial_i h)}
							\nonumber\\
			& =&\! \mathcal{N} \, \exp\!\bigg[\!-g_{ij}\,\Delta \zeta_i \Delta \zeta_j/(4D_0 \Delta t)\bigg],
\end{eqnarray}
where $G_0(x,y,z)=(4\pi D_0 \Delta t)^{-3/2} \exp [ - {\mathbf r}^2/4D_0 \Delta t ]$
is the usual Green's function for homogeneous 3D diffusion and the normalization is given
by $\mathcal{N}=\sqrt{g}/(4\pi D_0 \Delta t)$. Formally, 
this distribution defines a generalized  Wiener process \cite{Gardiner}, $\zeta_i(t)$, 
(with the non-constant kernel $g_{ij}$)  in the base plane, 
whose properties are well-defined in the infinitesimal time-step limit; in particular, one has
\begin{equation}
  {\mathrm{d}}\zeta_i\,{\mathrm{d}}\zeta_j = 2D_0 \,(\mathbf{g}^{-1})_{ij}\,\,{\mathrm{d}}t,
\end{equation}
and ${\mathrm{d}}\zeta_i \,{\mathrm{d}}t =0$. 

Rewriting Eq. (\ref{eq:dis_x_i}) for an infinitesimal random jump, one thus obtains 
\begin{equation}
   {\mathrm{d}} x_i = - D_0 \,\bigg[(\mathbf{g}^{-1})_{jk}\,\partial_j \partial_k h  \bigg]\,\frac{\partial_i h}{g} \,{\mathrm{d}}t  
			+ {\mathrm{d}}\zeta_i. 
		\label{eq:x_i}
\end{equation}
Note that the second-order term in random jump (coming from the projection displacement $\Delta_n$) 
leads to a {\em drift}, which is of the order $\sim{\mathrm{d}}t$, while the noise
term scales as ${\mathrm{d}}\zeta_i\sim {\mathrm{d}}t^{1/2}$ and all higher-order terms formally vanish.

\begin{figure}
\begin{center}
\includegraphics[angle=0,width=6.5cm]{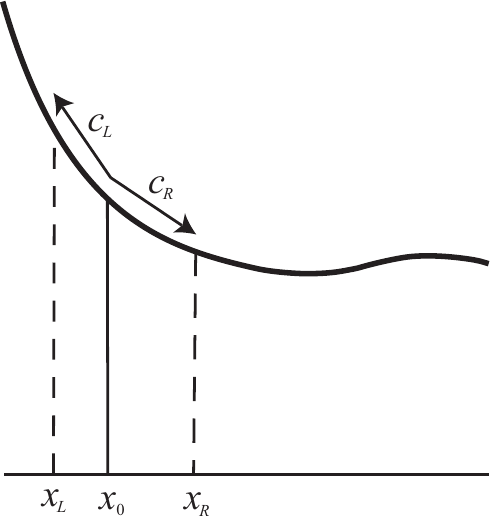}
\caption{Effect of curvature on the projected diffusive process: Curvature of a surface (here a line for simplicity)
 leads to deterministic drift within the 
Langevin equation.  Two random steps of equal magnitude ($c_L$ and $c_R$) cover the 
same curvilinear distance irrespective of direction, but  translate into different projected distances ($|x_L - x_0|$ 
and $|x_R - x_0|$) 
due to surface shape.  At a given position, the effect always favors displacement in the same direction, 
regardless of the exact (small) step size.  Note that both
slope and second derivative of the surface must be non-zero to break symmetry and drive the particle
in a specific direction. }
\label{fig:drift_illustration}
\end{center}
\end{figure}

The corresponding Langevin equation immediately follows as
\begin{equation}
   \dot{x}_i(t) = v_i + \tau_{ij}\, \eta_j(t), 
		\label{eq:Langevin}
\end{equation}
where the drift term, $v_i = v_i[h(\{x_k(t)\})]$, reads
\begin{equation}
   v_i = - D_0 \,\bigg[(\mathbf{g}^{-1})_{jk} \, \partial_j \partial_k h \bigg]\,\frac{ \partial_i h}{g} =  
		- 2 D_0 \,H\,\frac{ \partial_i h}{\sqrt{g}}. 
\label{eq:drift_h}
\end{equation}
It is induced by the mean curvature, $H$, and local gradient of the surface.  The fact
that surface curvature drives local drift is easily understood in the context of diffusion
on a 1D curved line (see Fig. \ref{fig:drift_illustration}).  In the presence of curvature and non-vanishing slope,
equidistant displacements in opposite directions along the curve necessarily lead to projected
displacements of different magnitude. 

In Eq. (\ref{eq:Langevin}),  we have defined $ \tau_{ij} \eta_j(t) = {\mathrm{d}}\zeta_i/{\mathrm{d}}t$,  where 
$\eta_i(t)$ is a Gaussian-distributed white noise of zero mean, $\langle \eta_i(t)\rangle =0$, and variance 
\begin{equation}
	\langle \eta_i(t)\, \eta_j(t')\rangle = 2D_0 \,\delta_{ij}\,\delta(t-t'),  
\end{equation}
and  $\tau_{ij} = \tau_{ij} \big[h(\{x_k(t)\})\big]$  is the square-root of the (positive-definite) inverse metric tensor, i.e. 
\begin{equation}
  (\mathbf{g}^{-1})_{ij} = \tau_{ik} \,\tau_{jk}. 
\end{equation}
The noise term in Eq. (\ref{eq:Langevin}) is thus multiplied by a factor containing explicit 
dependence on $(x,y)$, a clear manifestation of  ``multiplicative noise'' \cite{vankampen,Gardiner}.  
The derivation we have presented suggests that the noise must be treated within the Ito prescription \cite{vankampen,Gardiner}.
The noise prefactor may be calculated by diagonalization of the inverse metric tensor as  \cite{Note_diagon}
\begin{equation}
  \tau_{ij} = \delta_{ij} - \frac{\partial_i h\, \partial_j h}{g + \sqrt{g}}. 
  \label{eq:tau_ij}
\end{equation}
This completes the derivation of the Langevin equation for free diffusion on a curved surface
in Monge gauge. 

The projected motion of a particle confined to a Monge-gauge surface may be regarded as formally equivalent
to the motion of a particle in a 2D complex  inhomogeneous environment characterized by  an 
 effective ``potential", $U(\{x_i\})$, and a locally varying diffusion matrix, ${\mathcal D}_{ij}(\{x_k\})$, 
 both determined by the surface metric. The latter may be read off directly from the noise term as
\begin{equation}
 {\mathcal D}_{ij}=D_0 \, (\mathbf{g}^{-1})_{ij}. 
 \label{eq:D_ij}
\end{equation}
The explicit form of the effective potential may be seen more clearly from an equivalent 
Fokker-Planck equation \cite{Halle}. Using methods described in \cite{Risken,Gardiner,Zinn-Justin}, 
one can derive the following Fokker-Planck equation from the Langevin equation (\ref{eq:Langevin}),
i.e.  
\begin{eqnarray}
  \partial_t {\mathcal P}(\{x_i\}, t) &=& D_0 \,\partial_i\bigg[ \sqrt{g} \, (\mathbf{g}^{-1})_{ij}\, \partial_j \bigg(\frac{{\mathcal P}}{\sqrt{g}} \bigg)\bigg]\\
      &=&  \partial_i \bigg[ \partial_j\bigg({\mathcal D}_{ij} \, {\mathcal P}\bigg) - v_i \,{\mathcal P}\bigg]\\
  &=&  \partial_i\bigg[ {\mathcal D}_{ij} \bigg( \partial_j {\mathcal P} - u_j {\mathcal P}\bigg)\bigg],
  \label{eq:FP}
\end{eqnarray}
which governs the time evolution of the projected probability density, ${\mathcal P}(\{x_i\}, t)$, with the
normalization condition $\int {\mathrm{d}}^2 x\,  {\mathcal P}(\{x_i\}, t) = 1$. 
In the last equation, $u_j= - \partial_j U$ stands for the gradient of the effective dimensionless potential \cite{Halle}
\begin{equation}
   U(\{x_i\})= -\frac{1}{2}\ln g. 
   \label{eq:U}
\end{equation}



\section{Diffusion over a static surface}
\label{sec:quenched}

\subsection{Brownian Dynamics simulations on a quenched surface}
\label{subsec:sim_quenched}

In order to demonstrate how the projected diffusion coefficient, $D$, is affected by 
static surface undulations, 
we consider diffusion over periodic surfaces $h(\{x_i+n_iL\})=h(\{x_i\})$ with $i=1,\ldots, d$ in 
$d=1$ and 2 surface dimensions and $n_i$ being integers (hence, in 2D, the surface consists of  
identical square-like patches of size $L\times L$). 
This is a convenient choice for numerical simulations, and allows for comparison with existing works. 

The projected diffusion coefficient, $D$, is calculated by numerically evolving the projected 
position, $x_i(t)$, of a single particle (or equivalently, many non-interacting particles) 
according to the Langevin equation (\ref{eq:Langevin}) and then evaluating 
\begin{equation}
   D = \lim_{t\rightarrow\infty} \frac{1}{2d\,t}\sum_{i=1}^{d}\overline{[x_i(t)-x_i(0)]^2}, 
 \label{eq:Dsim_def}
\end{equation}
where the bar sign denotes average over an ensemble of particle trajectories with random
initial conditions, $x_i(0)$ (as well as  an ensemble of surface configurations in the case of the
random surfaces introduced in Section \ref{subsubsec:sim_quenched_memb}). 
In all cases considered here, we
choose surfaces (or ensembles of surfaces) with sufficient symmetry to insure that 
the principal directions $x$ and $y$ are equivalent. (The sum over $i$ in Eq. (\ref{eq:Dsim_def}) only serves 
to enhance sampling.)

In the simulations, we apply a standard iterative (Brownian Dynamics) algorithm \cite{McCammon}  by discretizing the Langevin 
equation (\ref{eq:Langevin}) using sufficiently small time steps of length $\delta t$.  
Hence at time $t=n\, \delta t$ (with integer $n\geq 0$), 
\begin{equation}
   x_i(n+1)=x_i(n) + v_i\!\big[x_k(n)\big] \delta t + \sqrt{2D_0 \delta t}\,  \tau_{ij}\!\big[x_k(n)\big] w_j(n), 
 \label{eq:langevin_discrete_res}
\end{equation}
where $w_i$ are  Gaussian-distributed random numbers with zero mean, $\langle w_i(n)\rangle =0$,
and unit variance $\langle w_i(n)\,  w_j(m)\rangle = \delta_{ij} \, \delta_{nm}$.  This form of the
discrete equation relies on the Ito interpretation of Eq. (\ref{eq:Langevin}) previously mentioned.
We emphasize that the lateral position of the particle is not constrained to any discrete lattice. 

Within our level of description, the projected diffusion coefficient, $D$, must be linearly proportional to 
$D_0$. This is because the only possible dimensional quantities affecting $D$ are $D_0$, $L$, and the
surface shape, $h(x, y)$, and the only
possible dimensionless quantity incorporating $D_0$ is the ratio $D/D_0$, which must  be expressible
in terms of the other dimensionless quantities in the physical problem \cite{dimensional-analysis}.  
That is, $D/D_0 = f\big[h(x,y)/L\big]$ is a functional of the reduced surface shape alone 
(see also Section \ref{subsubsec:reduced_rep}).  In the case of surfaces described 
by a single parameter, $A$ (such as the amplitude in the sinusoidal and cycloid surfaces specified by
Eqs. (\ref{eq:cos_2d}) and  (\ref{eq:cycloid}) below), $D/D_0$ is  a function only of $A/L$.


In the quenched simulations, we typically employed a rescaled time step  of 
$D_0\, \delta t/L^2 \sim10^{-6} - 10^{-5}$ and ran the simulations for approximately $10^6-10^7$ steps.  
It was verified that this choice of time step leads to convergence of the reported results for the 
roughest surfaces considered; i.e., further reduction of time step 
did not change the results within reported error-bars.
Statistical averages are made over an ensemble of $5\times10^3-10^4$  
trajectories  (and membrane configurations where necessary) as discussed above.
The error-bars are calculated using standard block-averaging methods \cite{block_av}. 
Note that for a typical  protein  in a lipid bilayer, the 
diffusion coefficient $D_0 \sim 1$~$\mu{\mathrm{m}}^2/{\mathrm{s}}$. Assuming a membrane patch of size $L\sim 0.1\mu$m, 
the simulation time steps fall in the range $\delta t\sim 0.01 - 1 \mu{\mathrm{s}}$. 
 
We remark that simulations may also be done with the random-walk algorithm 
defined via Eqs. (\ref{eq:RW_eq})-(\ref{eq:zeta_tangent})
as discussed in Ref. \cite{Holyst}. As we have shown in the previous Section, for sufficiently small time steps, 
this method is equivalent to the Langevin approach. The latter has the advantage that it implicitly
constrains the particles to the membrane. Furthermore, it is a 
dynamical equation that may be generalized to study more complicated systems with, for instance,
inter-particle or particle-membrane interactions. The random-walk algorithm is more convenient for 
surfaces with complicated geometry (such as 3D supramolecular assemblies)  
\cite{Holyst,Petersen}, where the present Monge-gauge Langevin equation is not 
adequate.  Also for surfaces with very large
gradients, it is clear that the Langevin approach will eventually require smaller time steps than the
3D random-walk approach and thus longer computational runs.  In practice, 
however, for the examples considered below, the Langevin simulations appear to be faster (typically by
about 50\% computer time in two dimensions) as comapred with the random-walk algorithm.  There
are simply less computations per time step in the Langevin approach due to the analytical reduction
of complexity when a Monge-gauge description is possible.  Random-walk simulations 
were carried out on a number of the reported test-cases for comparison purposes--always with
identical results to the Langevin simulations.  For simplicity, these random-walk results are explicitly 
displayed only in Figure \ref{fig:1d}.

We will first consider numerics on  
simple periodic surfaces in one dimension where analytical results are available.  We then progress
to simple 2D surfaces  (i.e., a sinusoidal surface and a crested cycloid  \cite{King})
and finally the case of an elastic membrane.

\subsection{One dimension: exact result and simulations}
\label{subsec:sim_quenched_1d}

In 1D, the local diffusion matrix (\ref{eq:D_ij}) is a scalar and reads ${\mathcal D}=D_0/g$. Highly steep surface segments 
with large slopes relative to the $x$-axis ($g\gg 1$) represent regions with small local diffusion 
coefficient, ${\mathcal D}/D_0\ll 1$, and large negative effective (dimensionless) potential, 
$-U\gg 1$,  Eq. (\ref{eq:U}). Regions of high slope may be regarded as 
trapping potential wells (see also Fig.  \ref{fig:MembProfiles} below).  Physically, the cause of this 
effect is clear.  In regions of high slope, a particle which travels a large distance along the contour of the curve travels a 
relatively much smaller projected distance.  The majority of travel is 
occurring in the $z$ direction, not contributing to the projected diffusion coefficient, $D$. 

\begin{figure}[t!]
\begin{center}
\includegraphics[angle=0,width=8.cm]{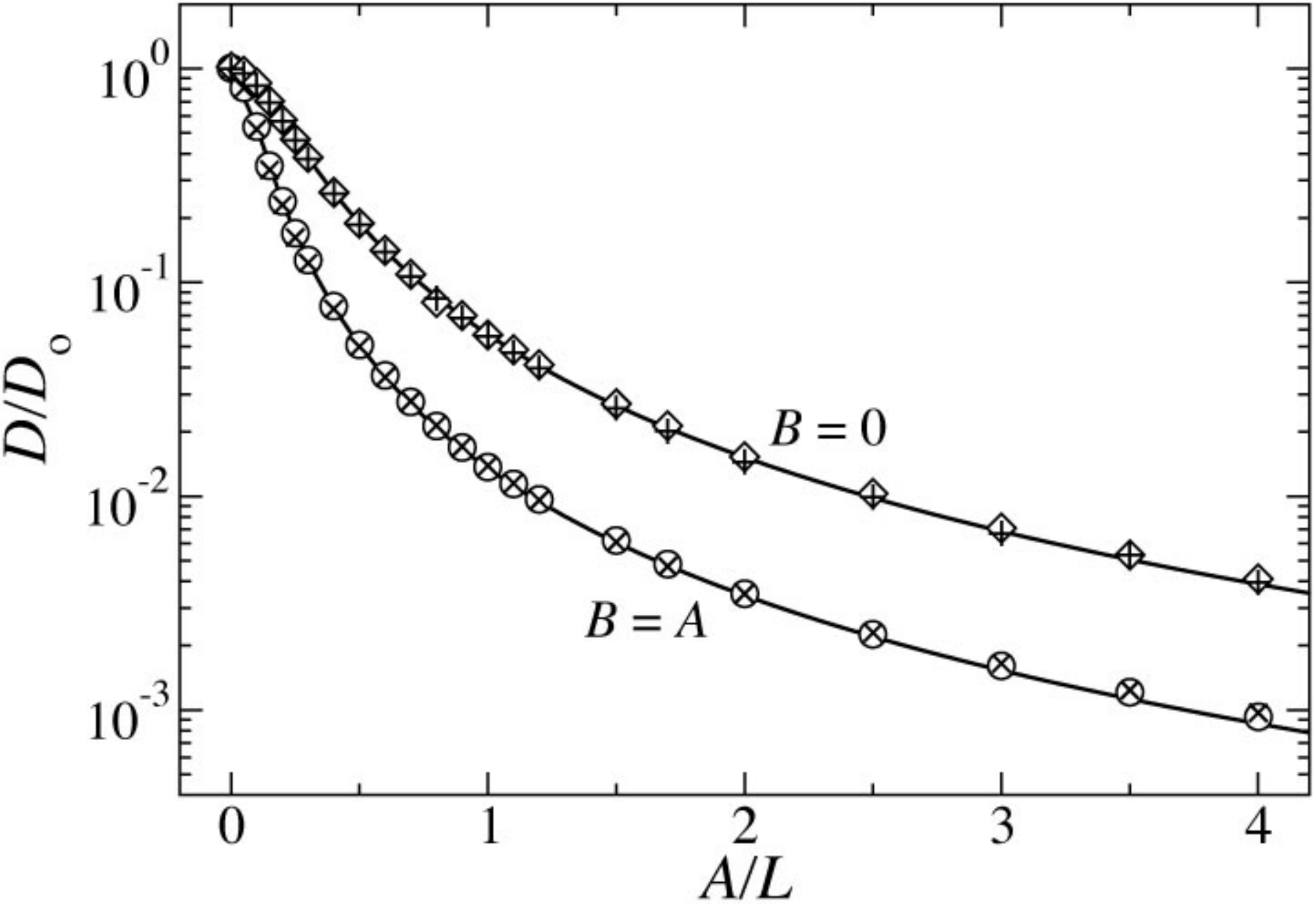}
\caption{Rescaled projected diffusion coefficient as a function of rescaled curve amplitude for free particle 
diffusion over a quenched 1D curve defined by Eq. (\ref{eq:cos_1d}) for two values of $B=0$ and $B=A$ as
indicated on the graph. Symbols show simulation data using Langevin dynamics (open diamonds and circles)
and random-walk algorithm \cite{Holyst} (cross and plus symbols). The solid curves show
the exact result, Eq. (\ref{eq:1d_exact}). Error-bars are smaller than the size of symbols.}
\label{fig:1d}
\end{center}
\end{figure}

The exact 1D expression for the projected diffusion coefficient along
an arbitrary periodic curve  may be written in terms
of the metric determinant, $g=1+(\partial_x h)^2$, as  \cite{Jackson,Festa,Halle_1d} 
\begin{equation}
  \frac{D}{D_0} =  \bigg(\frac{1}{L}\int_0^L\! {\mathrm{d}} x \, \sqrt{g} \bigg)^{-2} = \left\langle \frac{1}{\sqrt{g}} \right\rangle^2 = \left ( \frac{L}{L_c} \right)^2,
  \label{eq:1d_exact}
\end{equation}
where brackets denote a contour average  defined as 
$\langle \ldots \rangle = \int_0^L\! {\mathrm{d}} x \, \sqrt{g}\, (\ldots)/\int_0^L\! {\mathrm{d}} x \, \sqrt{g}$, 
 and $L_c$ is the contour length of the curve over a single projected period $L$. 

It is not surprising that analytical results are available in 1D.  The usual 1D Green's
function for diffusion predicts statistically Gaussian behavior for motion along the contour length 
of the curve.  For a Monge-gauge curve, this solution is readily transformed to the 
$x$-axis projected results, since contour length and projected length may be unambiguously mapped
to one another.   The projected solution in turn leads to Gaussian behavior in the long time limit with a 
simple rescaling by the square of the projected length to contour length ratio \cite{Halle_1d}.

In Figure \ref{fig:1d}, we plot the Langevin simulations results (open symbols) for a 
quenched periodic curve defined as
\begin{equation}
   h(x) = A \cos \frac{2\pi x}{L} + B \cos \frac{4\pi x}{L}. 
         \label{eq:cos_1d}
\end{equation}
for two different values of $B=0$ and $B=A$. 
As seen, the numerical  results exactly coincide with the analytical curve,  Eq. (\ref{eq:1d_exact}), 
confirming thus the validity of the present Langevin approach.  
For comparison, we have also plotted in the Figure  the simulation data  (cross and plus symbols) 
obtained by applying the random-walk algorithm \cite{Holyst},  Eqs. (\ref{eq:RW_eq})-(\ref{eq:zeta_tangent}), 
which, as expected, coincide with both Langevin simulations data and exact results.

\subsection{Diffusion on a 2D quenched manifold}
\label{subsec:sim_quenched_2d}

Unlike the 1D problem, a rigorous result for the projected diffusion coefficient is not
known for a quenched 2D surface. This relates to an intrinsic difference between 
Brownian motion on 1D and 2D manifolds \cite{Halle}. Namely, on a 1D manifold, the distance between
two points is given by the contour length which provides 
a natural parametrization, while on a 2D manifold, 
many different paths connect two given points, leading to 
analytical difficulties for establishing an exact result for a static surface \cite{Halle,Halle_1d}. 
Nonetheless, several approximate results have been presented in the past 
that we shall briefly review below following the discussions by Gustafsson et al.  \cite{Halle}.  

\subsubsection{Approximate analytical results}

First, one may establish an upper bound and a lower bound that limit the  
admissible values for $D$ using variational considerations \cite{Jackson,Halle}. 
The result for the upper bound in the quenched surface limit reads \cite{Halle}
\begin{equation}
\frac{D}{D_0}\bigg|_{\mathrm{upper}}= \frac{1}{2}\left(1+\left\langle \frac{1}{g} \right\rangle\right), 
\label{eq:upper}
\end{equation}
 where  brackets denote a surface average  defined as 
$\langle \ldots \rangle = \int_0^L\! {\mathrm{d}}^2 x \, \sqrt{g} \, (\ldots)/\int_0^L\! {\mathrm{d}}^2 x \, \sqrt{g}$.

\begin{figure}[t!]
\begin{center}
\includegraphics[angle=0,width=8.cm]{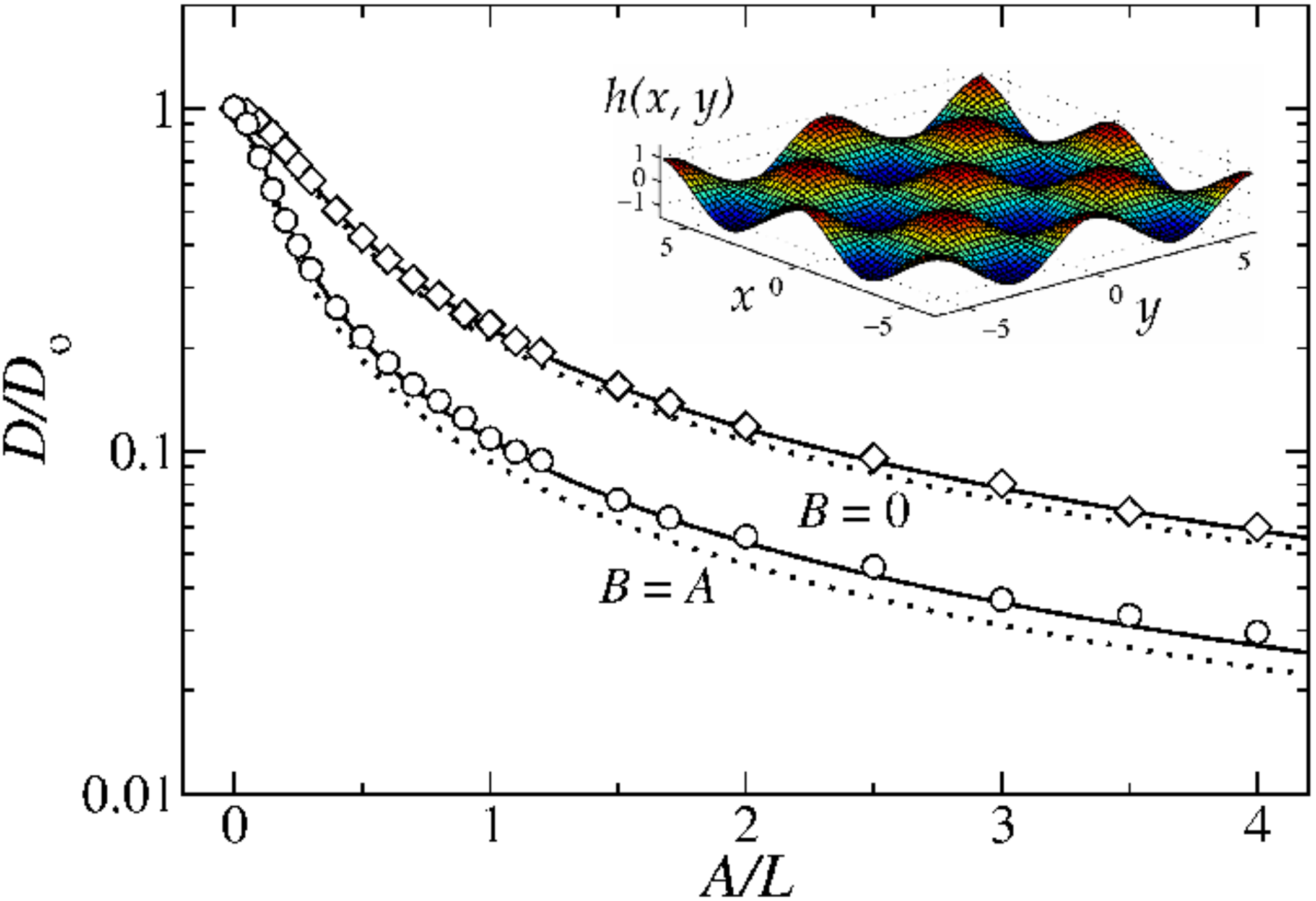}
\caption{Rescaled projected diffusion coefficient as a function of rescaled surface amplitude for free particle diffusion over
a quenched surface defined by Eq. (\ref{eq:cos_2d}) for two values of $B=0$ and $B=A$ as
indicated on the graph. Symbols show Langevin simulation data (open diamonds and circles), 
while solid and dotted curves show the area-scaling and the effective medium predictions, 
Eqs. (\ref{eq:area_scaling}) and (\ref{eq:2d_ema}), respectively. Inset shows surface profile 
for $B=0$, $A=1$ and $L=2\pi$.  Error-bars are smaller than or equal to symbol size. }
\label{fig:2d_two}
\end{center}
\end{figure}

 Intuitively, one expects that the quenched 1D result, Eq. (\ref{eq:1d_exact}), establishes a lower bound since 
correlation effects induced by quenched surface undulations are expected to be stronger in 1D \cite{Halle_1d}. 
Using variational methods, an expression similar to the 1D result may be obtained as  \cite{Halle}
\begin{equation}
\frac{D}{D_0}\bigg|_{\mathrm{lower}} = \frac{2\langle 1/\sqrt{g}\rangle^2}{1+\langle 1/g \rangle}. 
\label{eq:lower}
\end{equation}

In addition to these variational bounds, a mean-field 
approach--referred to as effective medium approximation (EMA)--has been introduced \cite{Halle,Solokov}, 
which is supposed to provide a good approximation 
for projected diffusion coefficients on strongly disordered quenched surfaces, i.e. 
\begin{equation}
\frac{D}{D_0}\bigg|_{\mathrm{EMA}} = \frac{1}{\langle \sqrt{g} \rangle}. 
\label{eq:2d_ema}
\end{equation}

Finally, another plausible approximation for  $D$ can be deduced based on naive scaling
arguments. On dimensional grounds, one might expect that an effective local isotropic 
projected diffusion coefficient would scale with the ratio of local projected area to 
actual curvilinear area, i.e., $D/D_0 \sim {\mathrm{d}}\mathcal{A}_\bot/{\mathrm{d}}\mathcal{A} = 1/\sqrt{g}$
(see Eq. (\ref{eq:diff_area})). 
Averaging this result over the entire surface leads to 
\begin{equation}
  \frac{D}{D_0}\bigg|_{\mathrm{area\,\, scaling}} = \left\langle \frac{1}{\sqrt{g}} \right\rangle 
  	= \frac{{\mathcal{A}}_\bot}{{\mathcal{A}}}
\label{eq:area_scaling}
\end{equation}
where the second equality follows from the definition of the averaging process as used in
Eq. (\ref{eq:upper}).  One could have also guessed the second equality by applying
similar arguments to the entire surface area instead of making local considerations.
This conjecture finds some support in the fact that it is exact in 1D (see Eq. (\ref{eq:1d_exact})). 
To our knowledge, no systematic derivation has been given so far in favor of the {\em area-scaling} prediction, 
though it has been discussed in several previous works \cite{Halle,Gov,King}.

\begin{figure}[t!]
\begin{center}
\includegraphics[angle=0,width=8.cm]{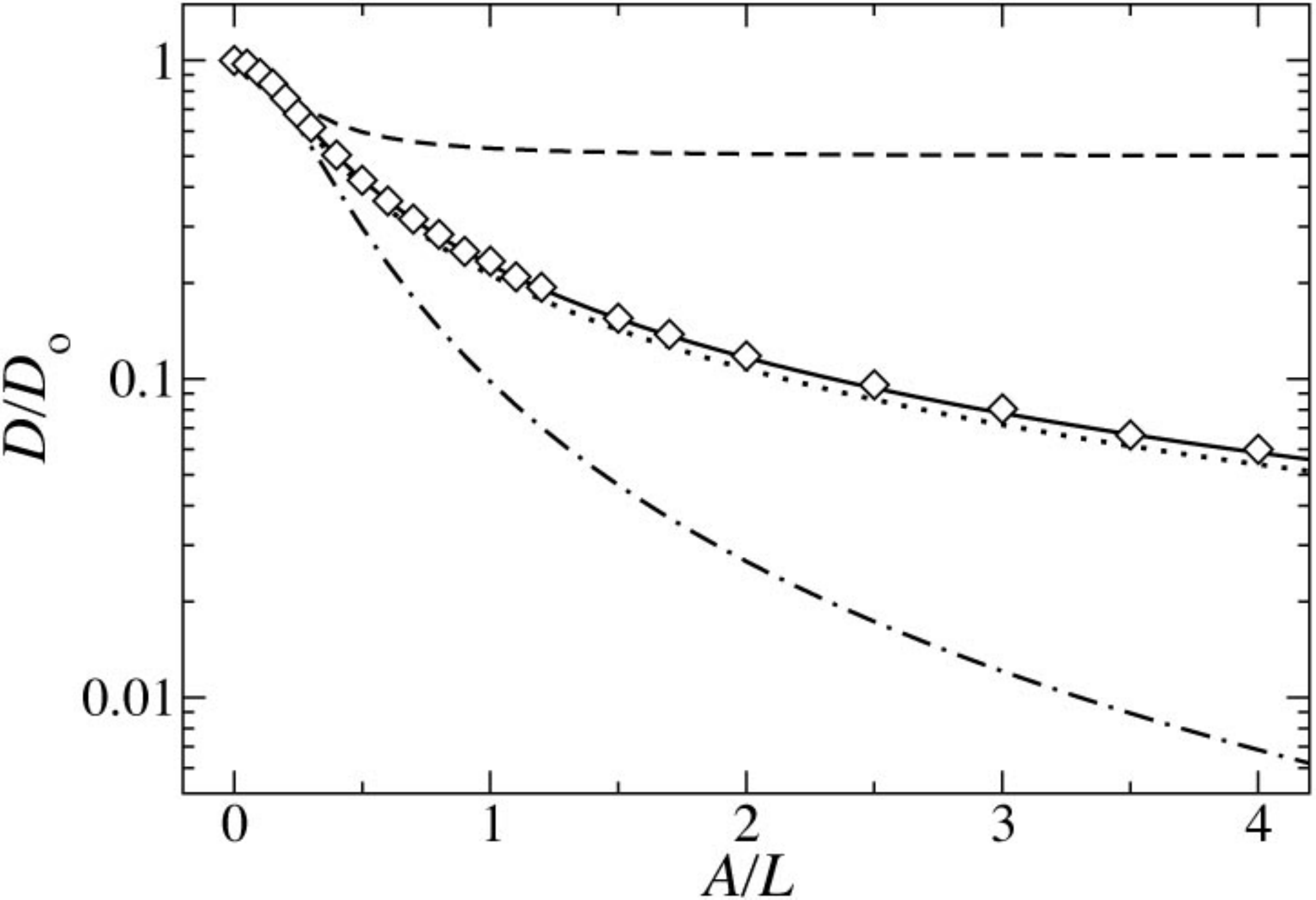}
\caption{ Same as Fig.  \ref{fig:2d_two} but for $B=0$. 
Symbols represent  simulation data, 
dashed and dot-dashed curves are the predicted upper and lower bounds, Eqs. (\ref{eq:upper}) and (\ref{eq:lower}), 
and solid and dotted curves show the area-scaling prediction, Eq. (\ref{eq:area_scaling}), and the effective medium prediction,
Eq. (\ref{eq:2d_ema}),  respectively.}
\label{fig:2d}
\end{center}
\end{figure}

\subsubsection{Simple periodic surfaces}

In order to examine the aforementioned analytical approximations, 
we first consider a simple periodic surface defined as 
\begin{equation}
h(x, y) = A \cos \frac{2\pi x}{L} \cos \frac{2\pi y}{L} + B \cos \frac{4\pi x}{L} \cos \frac{4\pi y}{L}. 
  \label{eq:cos_2d}
\end{equation}

Using Brownian Dynamics simulations as described in Section \ref{subsec:sim_quenched}, we calculate the projected 
diffusion coefficient for two different cases with $B=0$ and $B=A$ (shown by open symbols in Fig.  \ref{fig:2d_two}). 
As seen, in both cases the simulation data for $D/D_0$ as
a function of $A/L$ agree well with 
the area-scaling prediction (solid curves). There are strong deviations from both upper and lower bounds  
as shown in Figure  \ref{fig:2d} (compare with dashed and dot-dashed curves, respectively). 
But for the above surface, the area-scaling result stays very close to the effective medium 
prediction (dotted curve). 
Note that for small $A/L$, the surfaces are nearly flat and different approximations work similarly well
in this limit.   Only 
in the large $A/L$ limit, can one distinguish between different predictions. 
For  $A/L\gg 1$, the upper bound value for $D/D_0$ tends to 1/2, while the lower bound decays
as $\sim\! (A/L)^{-2}$. The simulation data (as well as area-scaling and EMA predictions) 
fall off as $\sim\! (A/L)^{-1}$. 

\begin{figure}[t!]
\begin{center}
\includegraphics[angle=0,width=8.cm]{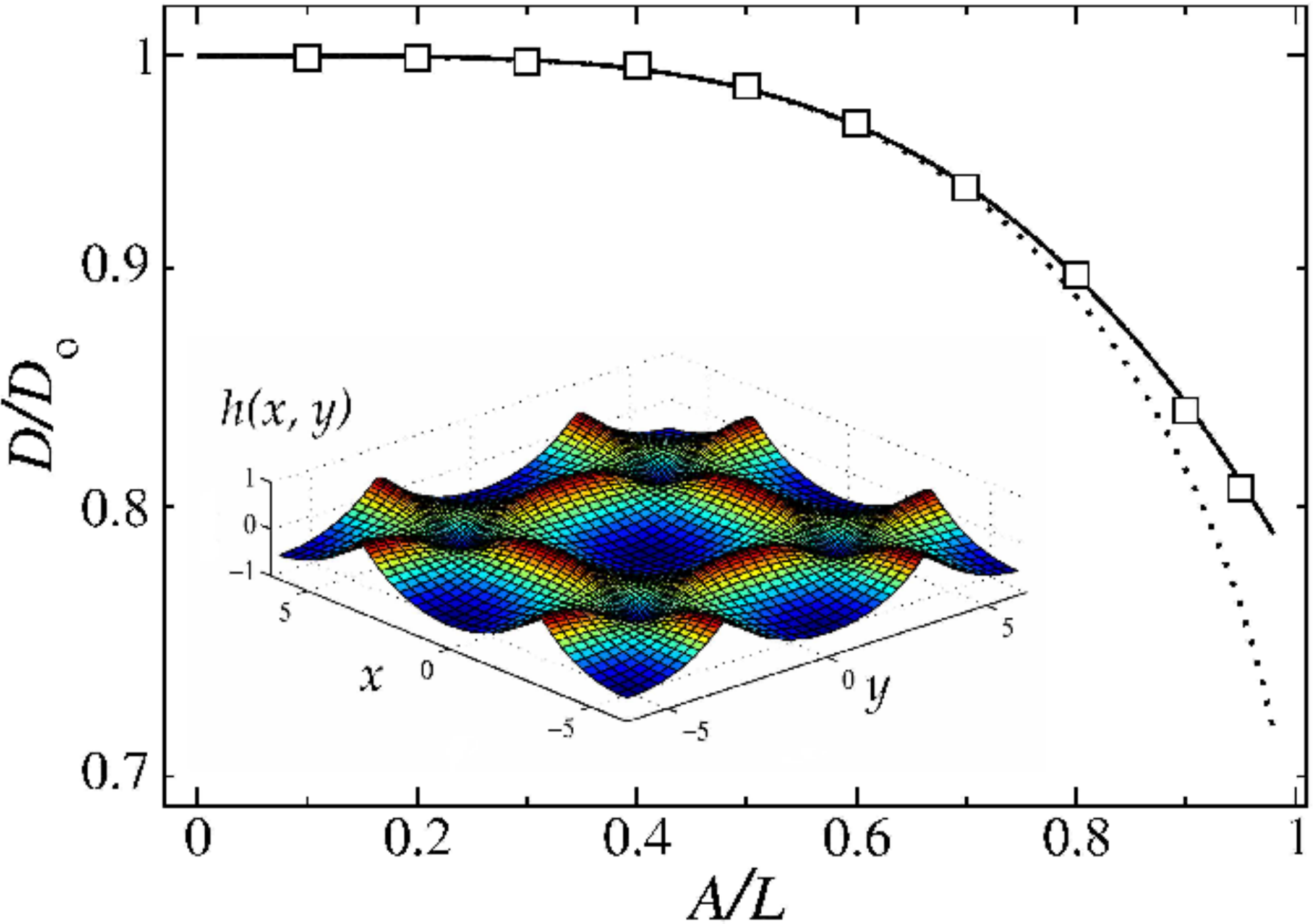}
\caption{Rescaled projected diffusion coefficient as a function of rescaled surface
amplitude for free particle diffusion over a quenched crested cycloid defined by Eq. (\ref{eq:cycloid}). 
Symbols show Langevin simulation data (open squares), 
and the solid and dotted curves show the area-scaling and the effective medium predictions, 
Eqs. (\ref{eq:area_scaling}) and (\ref{eq:2d_ema}), respectively. Inset shows the height profile
of a cycloid with $L=2\pi$ and $A=0.7$. Error-bars are smaller than or equal to symbol size.}
\label{fig:cycloid_D}
\end{center}
\end{figure}

Next we consider diffusion on a crested cycloid defined in a parametric form as
\begin{eqnarray}
x &= & u + \bigg(\frac{A}{2\pi}\bigg) \sin \left(\frac{2\pi u}{L}\right) \nonumber\\
y &= & v +  \bigg(\frac{A}{2\pi}\bigg)  \sin \left(\frac{2\pi v}{L}\right)  \nonumber\\ 
z &=& -\frac{A^2}{2\pi L} \,\cos \left(\frac{2\pi u}{L}\right)\, \cos \left(\frac{2\pi v}{L}\right),   
  \label{eq:cycloid}
\end{eqnarray}
where $z=h(x(u,v), y(u,v))$ represents the surface height (see Fig.  \ref{fig:cycloid_D} inset)
and $u$ and $v$ are real numbers.
 Note that the cycloid is not single-valued
 for $A/L > 1$ and at $A/L=1$, it shows singular cusp-like peaks.  Therefore, we restrict
 our discussion  to the range with $A/L<1$. 
Diffusion on this surface was previously investigated in Ref. \cite{King}, where 
it was suggested that the projected diffusion coefficient deviates from the area-scaling
prediction and follows the scaling law
 $D/D_0 = ({\mathcal A}_\bot/{\mathcal A})^{1.42}$. 

 In Figure \ref{fig:cycloid_D}, we have plotted the results from our simulations and compared
 with both the area-scaling and the effective medium approximation
 (solid and dotted curves, respectively). It is evident that the data closely follow the
 simple area-scaling rule $D/D_0 = ({\mathcal A}_\bot/{\mathcal A})$. 
 In the present case, the area-scaling and effective medium curves diverge for growing 
 $A/L$ ratio, with the latter systematically underestimating the projected diffusion coefficient.
 The spurious results presented in Ref. \cite{King} apparently stem from an incorrect treatment (neglect)
 of the surface metric within their simulation protocol.

\subsubsection{Quenched elastic membrane}
\label{subsubsec:sim_quenched_memb}

We now turn our attention to the case of a randomly undulated, quenched elastic membrane of size $L\times L$
with periodic boundary conditions.  In the continuum limit (large length scales), we use 
the Helfrich membrane model with  the Hamiltonian \cite{Safran}, 
\begin{eqnarray}
  {\mathcal H} &=& \frac{1}{2}\int_{{\mathcal A}_\bot}\! {\mathrm{d}}{\boldsymbol \rho}\, 
  					\left[K\,(\nabla^2_{\boldsymbol \rho}\, h)^2+\sigma\,(\nabla_{\boldsymbol \rho} \,h)^2\right] 
					\label{eq:H}\\
		& = &  \frac{1}{2L^2}\sum_{\mathbf{q}} \Omega_{\mathbf{q}}\, |h_{\mathbf{q}}|^2 \nonumber
\end{eqnarray}
where ${\boldsymbol \rho}=(x, y)$, $K$ is the bending rigidity, $\sigma$ the surface tension, and 
\begin{equation}
\Omega_{{\mathbf q}} = K q^4 + \sigma q^2, 
\label{eq:e_spectrum}
\end{equation}
the energy spectrum of membrane fluctuations. For simplicity,  we shall focus here only on two special cases, 
where only one of the elastic coefficients is non-zero. 

\begin{figure}[!t]
\begin{center}
\includegraphics[angle=0,width=8cm]{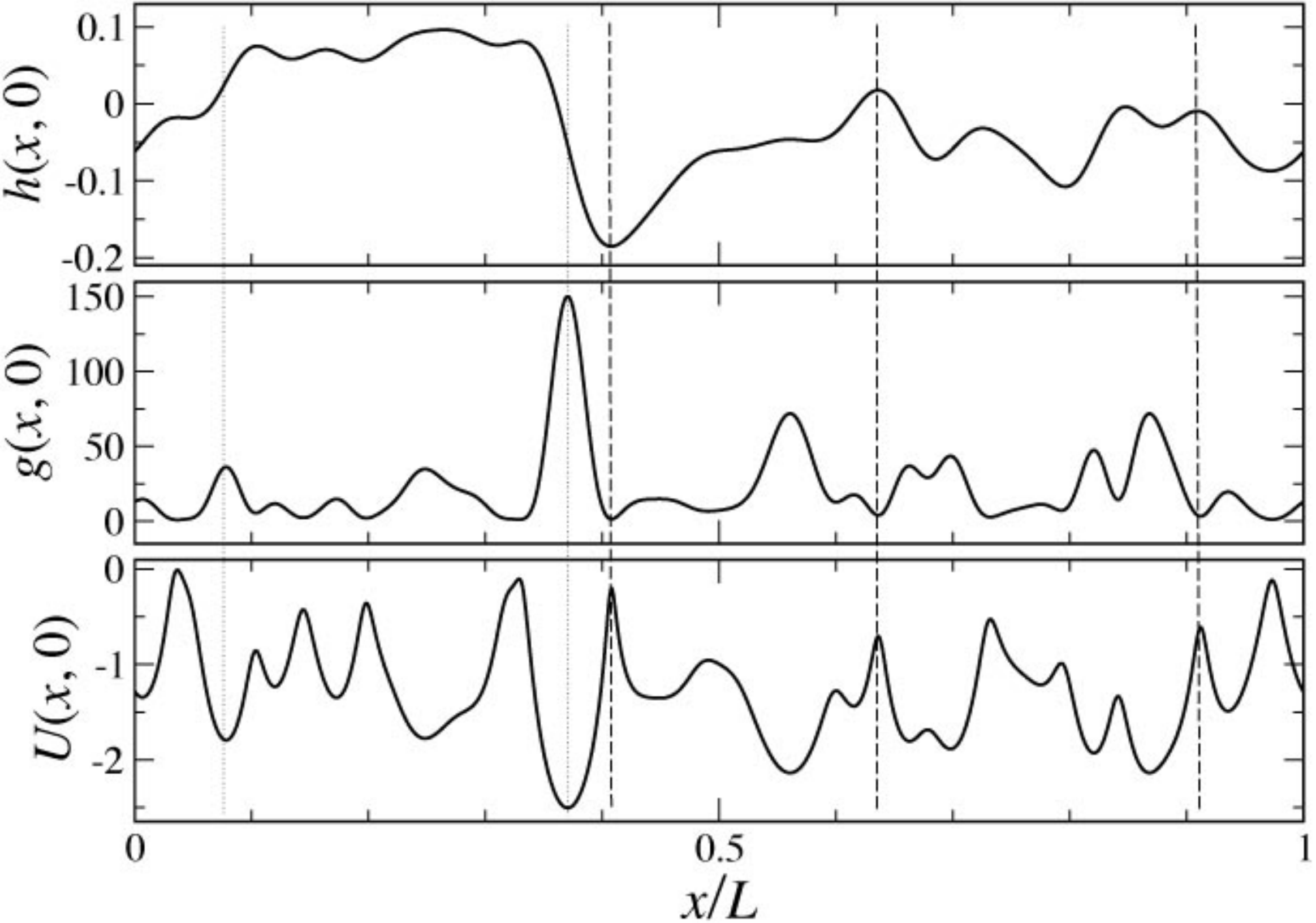}
\caption{Typical height profile, $h(x, y)$, metric determinant profile, $g(x, y)$, and
effective (dimensionless) potential profile, $U(x, y)$, for an elastic membrane with bending rigidity 
$K/(K_{\mathrm{B}}T)=0.01$ and surface tension $\sigma=0$ along the line $y=0$. 
Vertical lines show typical regions 
with minimum (thin lines) and maximum (thick lines) potentials.  
As discussed in the text, these curves
reflect a simple 1D slice through the surface.  Consequently, non-perfect correlations are observed
between, e.g.,  extrema in $h(x, 0)$, $g(x, 0)$ and $U(x, 0)$, 
reflecting the possibility of $y$-component contribution. }
\label{fig:MembProfiles}
\end{center}
\end{figure}

 At equilibrium, any configuration of the membrane may
be viewed as a linear combination of several Fourier modes, ${\mathbf q} = (q_x, q_y)$, 
similar to Eq. (\ref{eq:cos_2d}) but with 
amplitudes, $h_{\mathbf q}$, distributed according to the Boltzmann 
weight $\sim\! \exp(-{\mathcal H}/k_{\mathrm{B}}T)$. For numerical purposes, one needs to introduce a 
small-scale cut-off, $a$ (mimicking, for instance, a specific microscopic
length scale) in order to filter the irrelevant short-wave-length modes. The height profile is 
still continuous and defined in space via 
\begin{equation}
  h({\boldsymbol \rho}) = \frac{1}{L^2}\sum_{{\mathbf q}}   h_{{\mathbf q}}\, e^{{\mathrm{i}} {\mathbf q}\cdot{\boldsymbol \rho}}, 
 \end{equation}
where ${\mathbf q} = (q_x, q_y) = (2\pi n, 2\pi m)/L$ with $n, m$ being integral numbers in the range
$-M/2<n, m\leq M/2$ and $M=L/a$. 
In order to investigate
diffusion over elastic membrane surfaces, we first numerically generate many equilibrium 
membrane configurations consistent with the above Hamiltonian  by drawing $h_{\mathbf q}$
from normal distributions for each mode \cite{Brown,Lin2006} 
(see also Section \ref{sec:annealed_intermediate}).

\begin{figure}[!t]
\begin{center}
\includegraphics[angle=0,width=8cm]{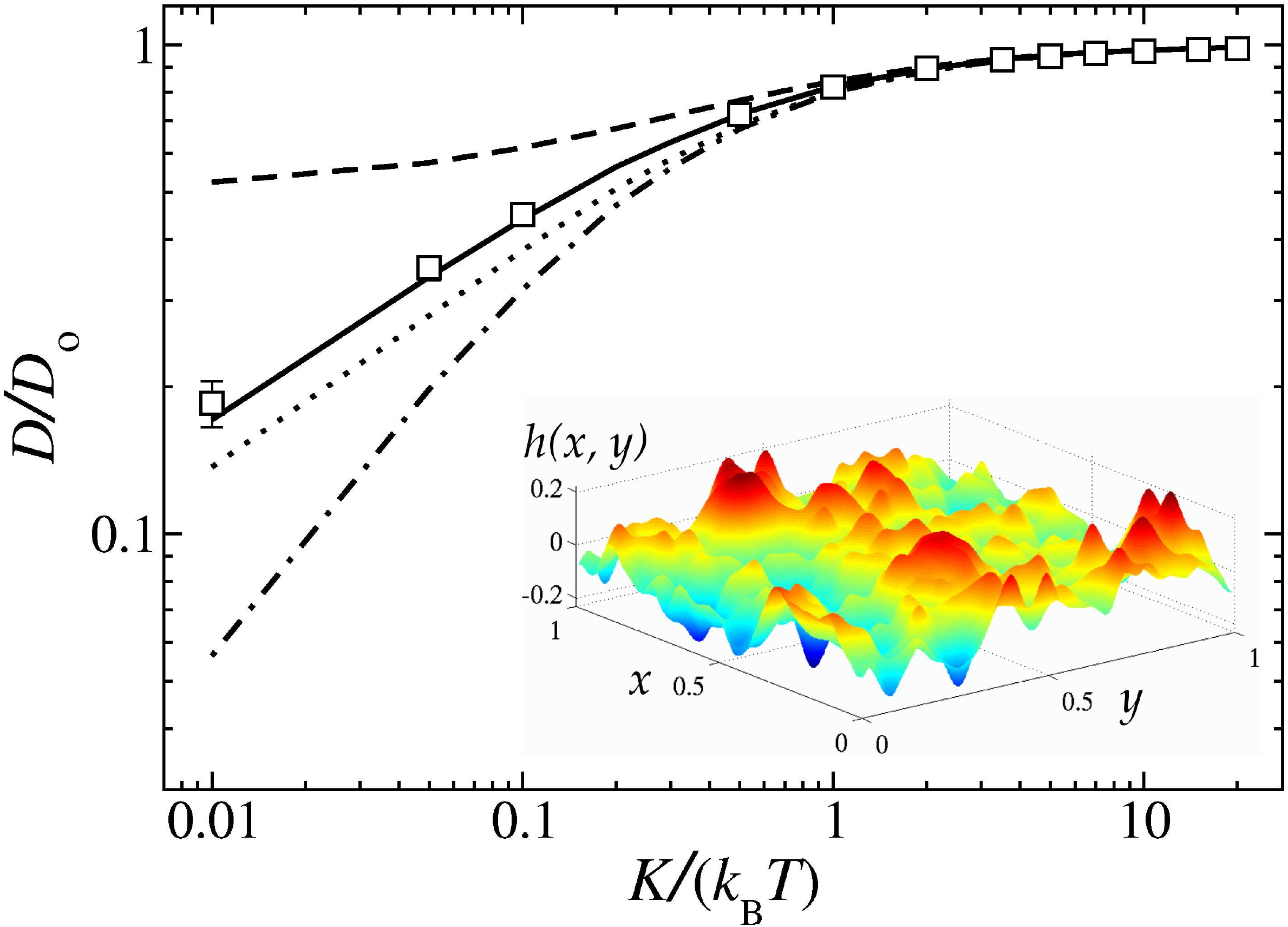}
\caption{Rescaled projected diffusion coefficient, $D/D_0$, of a freely diffusing particle on a quenched ruffled
membrane as a function of the rescaled membrane bending rigidity $K/(k_{\mathrm{B}}T)$ for $M=32$ and
surface tension $\sigma=0$.
Symbols represent simulation data, the dashed curve is the upper bound, Eq. (\ref{eq:upper}), the solid curve
is the area-scaling prediction, Eq. (\ref{eq:area_scaling}), the dotted curve is the effective
medium result, Eq. (\ref{eq:2d_ema}), and the dot-dashed curve
shows the predicted lower bound, Eq. (\ref{eq:lower}).   Inset shows a typical membrane configuration 
with  $K/(k_{\mathrm{B}}T)=0.01$, $L=1$ and $M=32$. Error-bars, if not shown, are smaller than or equal to the size of 
symbols. }
\label{fig:DvsK}
\end{center}
\end{figure}

\begin{figure}[!t]
\begin{center}
\includegraphics[angle=0,width=8cm]{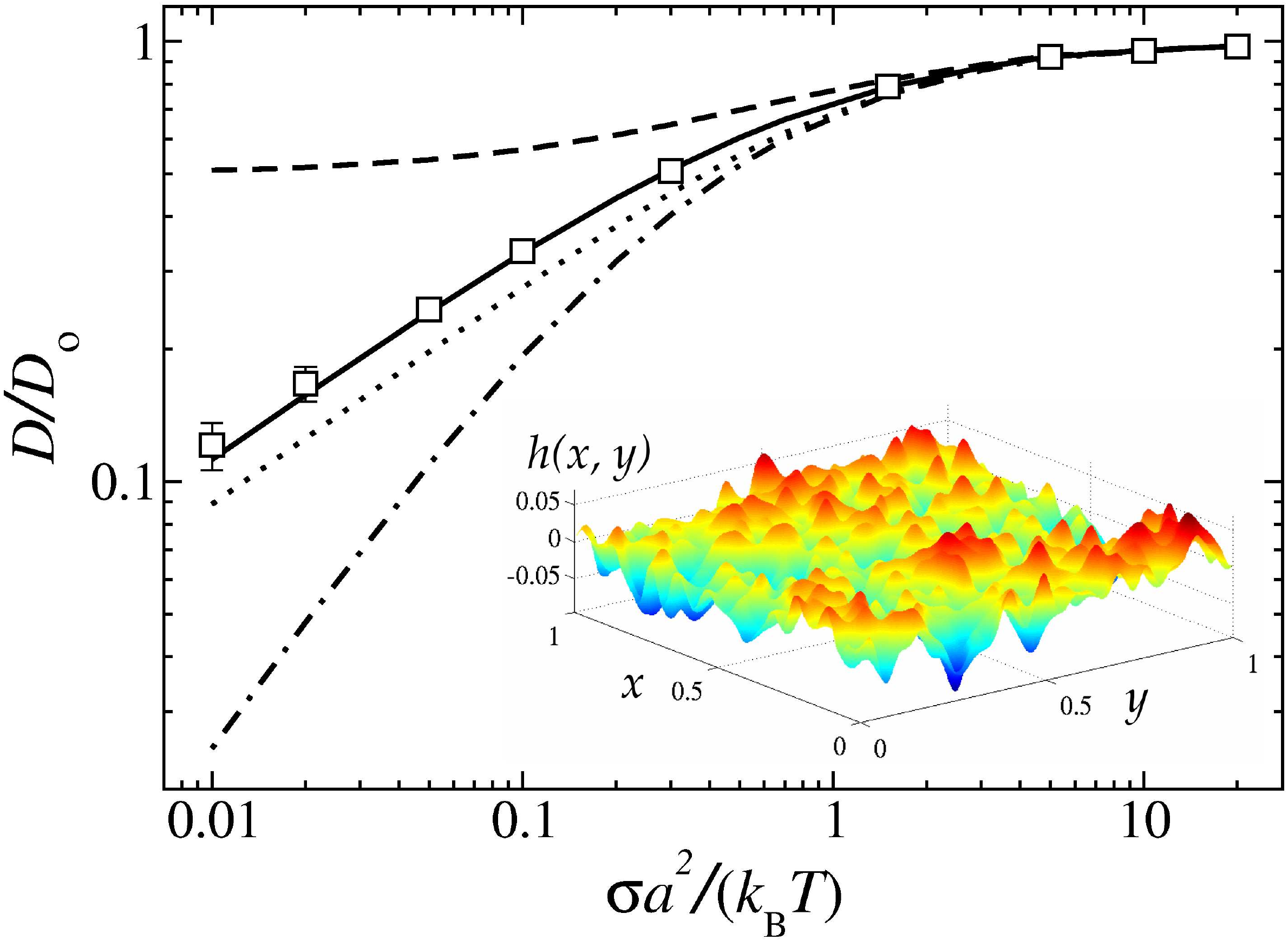}
\caption{Rescaled projected diffusion coefficient, $D/D_0$, of a freely diffusing particle on a quenched ruffled
membrane as a function of the rescaled membrane surface tension $\sigma a^2/(k_{\mathrm{B}}T)$ for $M=32$
and bending rigidity $K=0$.
Symbols represent simulation data, the dashed curve is the upper bound, Eq. (\ref{eq:upper}), the solid curve
is the area-scaling prediction, Eq. (\ref{eq:area_scaling}), the dotted curve is the effective
medium result, Eq. (\ref{eq:2d_ema}), and the dot-dashed curve
shows the predicted lower bound, Eq. (\ref{eq:lower}).   Inset shows a typical membrane configuration 
with  $\sigma a^2/(k_{\mathrm{B}}T)=0.2$, $L=1$ and $M=32$. Error-bars, if not shown, are smaller than or equal to the size of 
symbols.}
\label{fig:DvsSigma}
\end{center}
\end{figure}

We simulate the diffusion of a particle  using the method described in
Section \ref{subsec:sim_quenched}, and average the mean-square 
displacement over an ensemble of $5\times 10^3-10^4$ different equilibrium membrane configurations.
As previously discussed, the rescaled projected diffusion coefficient,  $D/D_0$, 
 is only a function of dimensionless membrane shape parameters, which, in the present case, are
the rescaled bending  rigidity $K/(k_{\mathrm{B}}T)$, rescaled surface tension $\sigma a^2/(k_{\mathrm{B}}T)$
and the ratio  $M=L/a$ (see also Section \ref{subsubsec:reduced_rep}). 
Here we choose $M=32$ for numerical convenience.  (This choice is most sensible assuming 
a membrane patch of size $L\simeq 100$~nm, which implies a small-scale cut-off of $a\simeq 3$~nm, 
comparable to the size of lipid molecules.) 
 
A typical membrane height profile along the $y=0$ axis is shown for $\sigma =0$ and $K/(k_{\mathrm{B}}T)=0.01$
in Figure \ref{fig:MembProfiles}  together with the profile of the surface metric determinant  and
the effective dimensionless potential, Eq. (\ref{eq:U}), experienced by a diffusing particle. 
Potential wells (peaks) correspond to steep (flat) segments of the membrane
and therefore a large (small) metric determinant as indicated by vertical lines on the graph. 
(Note that both $x$ and $y$ components of the
surface gradient contribute to $g(x, 0)$ and $U(x, 0)$, and that only the $x$-component 
can explicitly be seen in $h(x, 0)$ plot in the Figure.) 

The results for the disorder-averaged projected diffusion coefficient are  plotted in Figure \ref{fig:DvsK} 
as a function of 
the rescaled bending rigidity at fixed $\sigma=0$ (open symbols) and in Figure \ref{fig:DvsSigma}
as a function of the rescaled surface tension at fixed $K=0$. Both figures show qualitatively 
 similar behavior for the simulation data and theoretical predictions. (It is to be noted that the theoretical
 curves are computed using Eqs. (\ref{eq:upper})-(\ref{eq:area_scaling}) 
 and directly from the same ensemble of membrane configurations that are used in our Langevin simulations.) 
For large bending rigidities (or large surface tensions), the membrane stays close to a flat configuration and the data 
as well as all other theoretical predictions for $D/D_0$ tend to one (free planar diffusion)
in the same manner. The difference between these results becomes pronounced and they can be distinguished
in the limit of small bending rigidity (or small surface tension).  

Interestingly, we find that in both cases the simulation data agree well with the area-scaling 
prediction (\ref{eq:area_scaling}) (shown by a solid curve) 
down to very small values of $K$ and $\sigma$. In agreement with previous examples
studied in this paper, the effective medium approximation (dotted curve) diverges from the simulation data for small $K$ 
(or small $\sigma$) and, despite the suggestions made in Ref. \cite{Halle}, it represents a poorer approximation
 as compared to simple area scaling.



\section{Diffusion over a dynamically fluctuating surface}
\label{sec:annealed_intermediate}

So far we have only considered self-diffusion on a quenched surface. In this Section, we 
proceed by investigating dynamical effects from temporal surface fluctuations on the
particle diffusion over a two-dimensional Monge-gauge membrane. As it turns out, 
the effects of membrane fluctuations on the projected diffusion coefficient can vary considerably depending
on how fast membrane undulations relax relative to particle diffusion over the surface. 
In order to quantify this observation, one first needs to specify the dynamics of the surface. 

\subsection{Membrane dynamics}

We shall adopt the Helfrich membrane model described by a continuum time-dependent height profile, 
$h({\boldsymbol \rho}, t)$ with ${\boldsymbol \rho}=(x, y)$, 
whose equilibrium behavior may be determined by a Hamiltonian 
${\mathcal H}[h({\boldsymbol \rho})]$. 
As before, we assume periodic boundary conditions over a length scale of $L$ in lateral directions. 
The dynamics of membrane height fluctuations in a low-Reynolds-number hydrodynamic medium
is customarily described by a Langevin-type equation \cite{hohenberg,doi,lipowsky},  
\begin{equation}
 \frac{\partial h({\boldsymbol \rho}, t)}{\partial t}  = 
 		\int_{-\infty}^{+\infty}\!{\mathrm{d}}{\boldsymbol \rho}' \, \Lambda({\boldsymbol \rho}-{\boldsymbol \rho}') 
				\big[ F({\boldsymbol \rho}', t) + \eta({\boldsymbol \rho}', t)\big], 
				\label{eq:memb_langevin}
\end{equation}
where $\Lambda({\boldsymbol \rho}-{\boldsymbol \rho}')$ is the hydrodynamic interaction specified by 
the Oseen tensor (a no-slip boundary condition on the impermeable membrane surface is assumed) \cite{doi,Lin2006,Prost}, 
\begin{equation}
   \Lambda({\boldsymbol \rho}-{\boldsymbol \rho}') = \frac{1}{8\pi \eta \,|{\boldsymbol \rho}-{\boldsymbol \rho}'|}, 
\end{equation}
$F({\boldsymbol \rho}, t) $ is the force per unit area given by the functional derivative of the Hamiltonian, 
\begin{equation}
  F({\boldsymbol \rho}, t)  = -\frac{\delta {\mathcal H}}{\delta h({\boldsymbol \rho}, t)},
\label{eq:force_H}
\end{equation}
and $\eta({\boldsymbol \rho}, t)$ is a spatially-varying Gaussian, white random force of zero mean, 
$\langle \eta({\boldsymbol \rho}, t)\rangle =0$, that satisfies the fluctuation-dissipation theorem
\begin{equation}
  \langle \eta({\boldsymbol \rho}, t)\, \eta({\boldsymbol \rho}', t') \rangle = 
  			2 \,k_{\mathrm{B}}T\, \Lambda^{-1}({\boldsymbol \rho}-{\boldsymbol \rho}') \,\delta(t-t'). 
\end{equation} 
This choice of noise clearly ensures that the membrane relaxes to equilibrium at long times. 

To proceed, we express the membrane Langevin equation (\ref{eq:memb_langevin}) in terms of Fourier modes
$h_{{\mathbf q}}(t)$ with a discrete spectrum of wave-vectors ${\mathbf q} = (q_x, q_y) = (2\pi n, 2\pi m)/L$ where 
$-M/2<n, m\leq M/2$ and $M=L/a$ as discussed in Section \ref{subsubsec:sim_quenched_memb}. 
Hence, we have 
\begin{equation}
  h({\boldsymbol \rho}, t) = \frac{1}{L^2}\sum_{{\mathbf q}}   h_{{\mathbf q}}(t)\, e^{{\mathrm{i}} {\mathbf q}\cdot{\boldsymbol \rho}},
 \label{eq:fourier_t} 
 \end{equation}
using which we may write	 Eq. (\ref{eq:memb_langevin}) as  
\begin{equation}
\dot{h}_{{\mathbf q}}(t)  =  \Lambda_{\mathbf{q}}F_{{\mathbf q}}(t)
 				+ \sqrt{ k_{\mathrm{B}}TL^2\Lambda_{{\mathbf q}}}\, \,\xi_{{\mathbf q}}(t),
 		\label{eq:q_memb_langevin}
\end{equation}
where $F_{{\mathbf q}}(t)$ is the Fourier-transform of the force (\ref{eq:force_H}), $\Lambda_{{\mathbf q}}=1/(4\eta \,q)$
is that of the Oseen interaction and $\xi_{{\mathbf q}}(t)$ is a redefined Gaussian white noise with 
$\langle \xi_{{\mathbf q}}(t)\rangle =0$ and 
\begin{equation}
  \langle \xi_{{\mathbf q}}(t) \, \xi_{{\mathbf q'}}(t') \rangle = 
  			2\delta_{{\mathbf q}, -{\mathbf q}'} \,\delta(t-t'). 
\end{equation} 
It is understood that because $h({\boldsymbol \rho}, t)$ is real, $h_{{\mathbf q}}(t)$ is a complex number with 
$h_{{\mathbf q}}^\ast(t) = h_{-{\mathbf q}}(t)$, guaranteeing that the number of independent Fourier modes
remains $M^2$. 

Equation (\ref{eq:q_memb_langevin}) introduces an efficient algorithm, known as Fourier-Space 
Brownian Dynamics method \cite{Brown,Lin2006}, for simulating dynamics of biological membranes 
with arbitrary interaction potentials.

\subsection{Coupled membrane-particle equations}

 In what follows, we shall focus on the case of a  free elastic membrane 
described by the elastic Hamiltonian  (\ref{eq:H}).  
The force term in Eq. (\ref{eq:q_memb_langevin}) thus simplifies to a linear form 
$F_{{\mathbf q}}(t) =  - \Omega_{{\mathbf q}}\, h_{{\mathbf q}}(t)$, with $\Omega_{{\mathbf q}}$ defined 
in Eq. (\ref{eq:e_spectrum}). 

Putting together Eqs. (\ref{eq:Langevin}) and  (\ref{eq:q_memb_langevin}), we have the following set of 
dynamical equations for free particle diffusion on a free Monge-gauge elastic membrane,  
\begin{eqnarray}
   \dot{x}_i(t) &=& v_i+ \tau_{ij}\, \eta_j(t), 
		\label{eq:Langevin_particle}  \\
 \dot{h}_{{\mathbf q}}(t) &=& - \omega_{{\mathbf q}}\, h_{{\mathbf q}}(t)
  				+ \sqrt{ k_{\mathrm{B}}TL^2\Lambda_{{\mathbf q}}}\, \,\xi_{{\mathbf q}}(t),
 		\label{eq:q_memb_langevin_II}
\end{eqnarray}
where the decay rates, $\omega_{{\mathbf q}}$, are defined as  
\begin{equation}
\omega_{{\mathbf q}} = \Lambda_{{\mathbf q}}\, \Omega_{{\mathbf q}} = \Lambda_{{\mathbf q}}\,(K q^4 + \sigma q^2)
\end{equation}

It is to be noted that the particle dynamics (via Eq. (\ref{eq:Langevin_particle})) in the present description
is {\em passive} in the sense that it does not influence dynamics of the membrane (see the Discussion). 
However, as the height profile modes evolve in time (via Eq. (\ref{eq:q_memb_langevin_II})), they directly influence 
the curvature-induced drift, $v_i$, and the multiplicative noise factor, $\tau_{ij}$, which are related to the
local height of the membrane in real space, $h(x, y, t)$, as discussed in Section  \ref{sec:langevin}. 

The free membrane equations (\ref{eq:q_memb_langevin_II}) can be solved analytically as they 
correspond to a set of independent Ornstein-Uhlenbeck processes \cite{vankampen}. 
The time evolution of  $h_{{\mathbf q}}(t)$ (starting from the initial value $h^0_{{\mathbf q}} = h_{{\mathbf q}}(0)$),  
is governed by the conditional probability distribution 
\begin{eqnarray}
  P(h_{{\mathbf q}}, t| h^0_{{\mathbf q}}, 0) &=&\!\! \sqrt{ \frac{\beta  \Omega_{{\mathbf q}} }{2\pi L^2(1-e^{-2 \omega_{{\mathbf q}} t })  } }   \nonumber\\
  & \times&  \exp\bigg[ -\frac{\beta  \Omega_{{\mathbf q}} }{2L^2} 
  			 \frac{|h_{{\mathbf q}} - h^0_{{\mathbf q}}\, e^{-\omega_{{\mathbf q}} t }|^2}{1-e^{-2\omega_{{\mathbf q}} t }} \bigg]. 
\label{eq:cond_prob}
\end{eqnarray}

\subsection{Crossover from quenched to annealed disorder limit}

In general, the coupled dynamical equations (\ref{eq:Langevin_particle}) and (\ref{eq:q_memb_langevin_II}) can be used 
to study the diffusion process on a surface that may be static, as studied in previous Sections, or dynamic. 
The quenched disorder as well as the annealed disorder limits can thus be realized 
as special cases of Eqs.  (\ref{eq:Langevin_particle}) and (\ref{eq:q_memb_langevin_II}),
 by choosing the membrane-related parameters (such as bending rigidity or medium viscosity) 
such that the membrane decay rates, $\omega_{{\mathbf q}}$, tend to zero (static surface) or infinity (fast-fluctuating surface). Physically, however, these  same limits can also be achieved by taking a very large or small curvilinear particle diffusion coefficient, $D_0$. 
The coupling between particle and membrane motions introduces a new dimensionless parameter to the problem
not present in the quenched case (or the annealed case).  This parameter can be used to track the transition from quenched to annealed limits. 

\subsubsection{Reduced representation}
\label{subsubsec:reduced_rep}

In order to demonstrate this feature, we shall use a reduced (dimensionless) representation by rescaling length and time scales 
by a characteristic length and time, that is, using $\tilde x = x/L$, or equivalently, $\tilde {\mathbf q} = {\mathbf q}L$, 
$\tilde t = D_0t/L^2$, as well as $\tilde h(\tilde x, \tilde y, \tilde t)=h(x, y, t)/L$, or equivalently, 
$\tilde h_{\tilde {\mathbf q}}(\tilde t)=h_{{\mathbf q}}(t)/L^3$. Hence we obtain 
\begin{eqnarray}
   \dot{\tilde x}_i(\tilde t) &=& \tilde v_i+ \tilde \tau_{ij}\, \tilde \eta_j(\tilde t), 
		\label{eq:Langevin_particle_res}  \\
 \dot{\tilde h}_{\tilde {\mathbf q}}(\tilde t) &=& - \chi\, \tilde \omega_{\tilde {\mathbf q}} \, \tilde h_{\tilde {\mathbf q}}(t) 
 				+ \sqrt{\chi \, \tilde \Lambda_{\tilde {\mathbf q}}}\, \,\tilde \xi_{\tilde {\mathbf q}}(\tilde t),
 		\label{eq:q_memb_langevin_II_res}
\end{eqnarray}
where the rescaled functions, namely, rescaled drift, $ \tilde v_i $, rescaled multiplicative noise factor, $ \tilde \tau_{ij}$, 
as well as rescaled noise terms $\tilde \eta_i(\tilde t)$ and $\tilde \xi_{\tilde {\mathbf q}}(\tilde t)$,  
whose explicit forms are given in Appendix \ref{app:rescaled}, are independent of system parameters. The rescaled 
decay rate reads $\tilde \omega_{\tilde {\mathbf q}} = \tilde \Lambda_{\tilde {\mathbf q}}\, \tilde \Omega_{\tilde {\mathbf q}}$
where $\tilde \Omega_{\tilde {\mathbf q}} = \tilde K \tilde q^4 + \tilde \sigma \tilde q^2$ and
$\tilde \Lambda_{\tilde {\mathbf q}}=1/\tilde q$. 

Note that all physical system parameters have been condensed into four dimensionless
combinations. These are the 
rescaled bending rigidity $\tilde K=K/(k_{\mathrm{B}}T)$, rescaled surface tension 
$\tilde \sigma = \sigma L^2/(k_{\mathrm{B}}T)$, $M=L/a$, and an additional parameter
\begin{equation}
  \chi = \frac{k_{\mathrm{B}}T}{4\eta D_0 L}, 
\end{equation}
which we shall refer to as the {\em dynamical coupling parameter} as  it embeds in itself  
the dynamical  coupling between equations (\ref{eq:Langevin_particle}) and (\ref{eq:q_memb_langevin_II}). 
Indeed, for vanishing dynamical coupling parameter $\chi\rightarrow 0$, the membrane equation 
reduces to $\dot{h}_{{\mathbf q}}(t) =0$, representing a static surface.
 
One may also examine the conditional probability  in rescaled units 
\begin{eqnarray}
  \tilde P(\tilde h_{\tilde {\mathbf q}}, \tilde t| \tilde h^0_{\tilde {\mathbf q}}, 0) &=&\!\! 
  		\sqrt{ \frac{ \tilde \Omega_{\tilde {\mathbf q}} }{2\pi (1-e^{-2 \chi \tilde \omega_{\tilde {\mathbf q}} \tilde t })  } }   \nonumber\\
  & \times&  \exp\bigg[ -\frac{ \tilde \Omega_{\tilde {\mathbf q}} }{2} 
  			 \frac{|\tilde h_{\tilde {\mathbf q}} - \tilde h^0_{\tilde {\mathbf q}}\, e^{-\chi \tilde \omega_{\tilde {\mathbf q}} \tilde t }|^2 }
			 		{1-e^{-2\chi \tilde \omega_{\tilde {\mathbf q}} \tilde t }} \bigg], 
\label{eq:cond_prob_res}
\end{eqnarray}
which is defined via $\tilde P(\tilde h_{\tilde {\mathbf q}}, \tilde t| \tilde h^0_{\tilde {\mathbf q}}, 0) =P(h_{{\mathbf q}}, t| h^0_{{\mathbf q}}, 0)\,L^3$. 
For $\chi\rightarrow 0$, the static surface solution is recovered, that is
$\tilde P(\tilde h_{\tilde {\mathbf q}}, \tilde t| \tilde h^0_{\tilde {\mathbf q}}, 0) \rightarrow \delta(\tilde h_{\tilde {\mathbf q}}-\tilde h^0_{\tilde {\mathbf q}})$, 
while for  $\chi\rightarrow \infty$, it rapidly decays to the equilibrium distribution 
\begin{equation}
\tilde P(\tilde h_{\tilde {\mathbf q}}, \tilde t| \tilde h^0_{\tilde {\mathbf q}}, 0) \rightarrow \tilde P_{\mathrm{eq}}(\tilde h_{\tilde {\mathbf q}})
			= \bigg( \frac{\tilde \Omega_{\tilde {\mathbf q}}}{2\pi} \bigg)^{1/2} 
					e^{-  \tilde \Omega_{\tilde {\mathbf q}} |\tilde h_{\tilde {\mathbf q}}|^2/2}.
\label{eq:P_memb_eq}
\end{equation}
These two limits therefore reflect the 
 static surface (quenched disorder) and the fast-fluctuating surface (annealed disorder) limits, respectively. 
 Solving the set of equations  (\ref{eq:Langevin_particle_res}) and (\ref{eq:q_memb_langevin_II_res}) for a finite 
 intermediate value of $\chi$ allows one to investigate the {\em intermediate regime}, as we shall do later.

The reduced representation explicitly demonstrates
that any dimensionless quantity obtained via simulations must depend on individual system parameters {\em only} 
through the dimensionless combinations 
$\tilde K$, $\tilde \sigma$, $\chi$, and  $M=L/a$. 
This implies that the ratio of the projected diffusion coefficient 
and the curvilinear diffusion coefficient, $D/D_0$, has the general form
\begin{equation}
  \frac{D}{D_0} = f\big(\tilde K, \tilde \sigma, \chi, M\big),
 \label{eq:D_rescaledform}
\end{equation}
which could be surmised simply from dimensional considerations \cite{dimensional-analysis}.  
From a practical perspective, the reduced approach simplifies our investigation of 
the dependence of $D/D_0$ on various system parameters.  The addition of membrane motion
adds only a single dimensionless parameter beyond those necessary in describing the quenched
limit.  This parameter, $\chi$, may be thought of as a rescaled solvent viscoscity.  Since $\chi$
 appears only when surface dynamics is important, we tend to focus attention on
it when calculating $D/D_0$ over fluctuating surfaces.  However, we stress that all 
the dimensionless parameters of the system influence the value of $D/D_0$.

Another advantage of the reduced representation is that it immediately suggests a
``small parameter'' for use in a purturbative expansion around the analytical annealed limit
results.  In Appendix \ref{app:AEFV}, we present an operator expansion in powers of 
$\chi^{-1}$.  At zeroth order, the pre-averaging Fokker-Planck equation of Reister
and Seifert  \cite{Seifert} is recovered.  Higher-order corrections are  predicted,
but not pursued in this work.

\subsubsection{Characteristic time scales}
\label{subsubsec:time_scales}

The physical meaning of the dynamical coupling parameter, $\chi$, may be appreciated further when
looking at the typical time scales that characterize the dynamics of particle diffusion and that of 
membrane relaxation. 

The typical diffusion time over a length of size $q^{-1}$ may be written as 
$\tau^d_{{\mathbf q}} = 1/(D_0 q^2)$, while the relaxation time of an undulation of 
wave-length $q^{-1}$ is given by $\tau^m_{{\mathbf q}} = \omega_{{\mathbf q}}^{-1}$. 
Using the definition of the decay rate, $\omega_{{\mathbf q}}$, we have the time scale ratio
\begin{equation}
     \frac{\tau^d_{{\mathbf q}}}{\tau^m_{{\mathbf q}}} = \frac{K q^2 + \sigma }{4\eta D_0 q} 
     				=  \chi \bigg(\frac{\tilde K \tilde q^2 + \tilde \sigma }{\tilde q}\bigg). 
\label{eq:tau_ratio}
\end{equation}
In general, one can define the quenched disorder limit as the limit where the membrane 
relaxation time becomes much  larger than the diffusion time. While, the converse limit of vanishingly 
small relaxation time represents the annealed disorder limit. 
When the two time scales are comparable, one encounters an intermediate situation. 

Clearly, at short wave-lengths $q\rightarrow \infty$, one always deals with an annealed situation. 
But it is the long wave-length, $q\rightarrow 0$, behavior which is of primary practical importance 
in the context of physical measurements.   A membrane with a finite surface tension always exhibits 
annealed behavior at sufficiently long wave-lengths as seen from Eq. (\ref{eq:tau_ratio}), 
whereas a membrane with only bending rigidity exhibits quenched behavior at long wave-lengths.

It is evident that in a {\em finite} or {\em periodic} system, such as the ones
considered in this paper where the wave-vector is bounded from below by the surface periodicity $q>2\pi/L$, 
the transition from one limit to another can be established theoretically by changing the dynamical
coupling parameter over the range $0\leq \chi <\infty$ at fixed bending rigidity, surface tension and
$M$. This is 
indicated by the second equality in Eq. (\ref{eq:tau_ratio}). 
In what follows, we shall focus only on the case with zero surface tension, $\sigma =0$, 
where quenched behavior can be achieved within reasonable values for system parameters.

\subsection{Coupled particle-membrane simulation method}

In order to simulate diffusion on a dynamic surface
in an arbitrary dynamical situation, we apply the standard Brownian Dynamics algorithm by discretizing the time 
into steps of sufficiently small length similar to what we described in Section \ref{subsec:sim_quenched}. 
 In order to bring up the role of the dynamical coupling and
provide a unified analysis of the quenched-annealed crossover behavior, it is most convenient 
to simulate the system in the reduced 
representation using Eqs. (\ref{eq:Langevin_particle_res}) and (\ref{eq:q_memb_langevin_II_res}), 
where all various system parameters condense into a few relevant dimensionless parameters. 

We evolve particle position using small rescaled time steps  $\delta \tilde t$ such that 
at time $\tilde t=n\, \delta \tilde t$ (with $n\geq 0$ being an integer)
\begin{equation}
   \tilde x_i(n+1)=\tilde x_i(n) + \tilde v_i(n)\, \delta \tilde t + \sqrt{2 \delta \tilde t}\,  \tilde \tau_{ij}(n) \, w_j(n)
\label{eq:langevin_discrete_res_II}
 \end{equation}
for $i, j=1,2$, where $w_i$ are  Gaussian-distributed random numbers with zero mean, $\langle w_i(n)\rangle =0$, 
and unit variance $\langle w_i(n)\,  w_j(m)\rangle = \delta_{ij} \, \delta_{nm}$. Other rescaled functions appearing 
in the above equation are given explicitly in Appendix \ref{app:rescaled}. They are evaluated from
the instantaneous membrane height  $\tilde h\big(\{\tilde x_i(n)\}, n \big)=\tilde h\big(\tilde x(n), \tilde y(n), n\big)$ obtained 
from the Fourier modes $\tilde h_{\tilde {\mathbf q}}(n)$, via Eq. (\ref{eq:fourier_t}).

Note that to evaluate $\tilde h_{\tilde {\mathbf q}}$ at each time step, one can either use the discretized form 
of Eq. (\ref{eq:q_memb_langevin_II_res}), i.e. 
 \begin{equation}
   \tilde h_{\tilde {\mathbf q}}(n+1) = \tilde h_{\tilde {\mathbf q}}(n) 
   			- \chi \,\tilde \omega_{\tilde {\mathbf q}}\,  \tilde h_{\tilde {\mathbf q}}(n)\, \delta \tilde t
 				+ \sqrt{\chi  \tilde \Lambda_{\tilde {\mathbf q}}\,  \delta \tilde  t}\, \,\tilde R_{\tilde {\mathbf q}}(n),
				\label{eq:langevin_discrete_memb}
\end{equation}
(with random complex variables $\tilde R_{\tilde {\mathbf q}}$ of zero mean and variance 
$\langle \tilde R_{\tilde {\mathbf q}}(n) \, \tilde R_{\tilde {\mathbf q}'}(m)\rangle  = 2 \delta_{\tilde {\mathbf q}, -\tilde {\mathbf q}'}\, \delta_{nm}$),
or  directly draw $\tilde h_{\tilde {\mathbf q}}(n)$ values from the Ornstein-Uhlenbeck 
Gaussian distribution, Eq.  (\ref{eq:cond_prob_res}). 
The second method has the clear advantage that it is based on an exact result and is 
computationally easier to implement 
as previously performed in Ref. \cite{Brown}. 
We have used both methods in our simulations with the same results obtained 
within the statistical error-bars. 

The time step $\delta \tilde t$ must be chosen such that it ensures the convergence of 
Eq. (\ref{eq:langevin_discrete_res_II}) (and Eq. (\ref{eq:langevin_discrete_memb}), if applied).  
Even if the membrane dynamics are drawn from the exact distribution (\ref{eq:cond_prob_res}), 
care must still be taken to insure that a small enough $\delta \tilde t$ is used to properly 
capture the effect of membrane dynamics in Eq. (\ref{eq:langevin_discrete_res_II}).  Both
the drift and diffusion tensor for the particle depend upon the local height of the membrane and
it is to be expected that at large $\chi$, one needs to use 
time steps limited by the membrane dynamics as opposed to particle diffusion.  We typically
 choose rescaled time steps in the range $\delta \tilde t = 10^{-8} - 10^{-6}$, 
 and run the simulations for $10^5-10^7$ time steps.  (As indicated previously, step size is
 verified by checking for convergence of results.)
In the simulations, we measure the projected diffusion coefficient via the mean square displacement of 
a Brownian particle as defined in Eq. (\ref{eq:Dsim_def}).  
(Simulated mean square displacements converge more quickly with time to a 
diffusive long-time behavior for large $\chi$ or highly fluctuating membranes.)  
 We average the results  over an ensemble of $5\times10^3-10^4$  
particle trajectories  with random initial conditions for the particle as well as the membrane height profile. 

In our simulations, the coupling parameter is varied over the range $\chi = 10^{-4}-10^2$ 
and the rescaled membrane bending 
rigidity in the range $\tilde K = 0.01 - 10$,  with $M=L/a = 32$ 
being chosen as in the quenched  study in Section \ref{subsubsec:sim_quenched_memb}. 
Recall that at fixed bending rigidity and system size, changing $\chi = k_{\mathrm{B}}T/(4\eta D_0 L)$ 
may be viewed as changing either the curvilinear diffusion coefficient, $D_0$, or the medium viscosity, $\eta$, 
or a combination of both. They all have the same quantitative effect on $D/D_0$, once $\tilde K$ and
$M$ are fixed. 

\begin{figure}[!t]
\begin{center}
\includegraphics[angle=0,width=8cm]{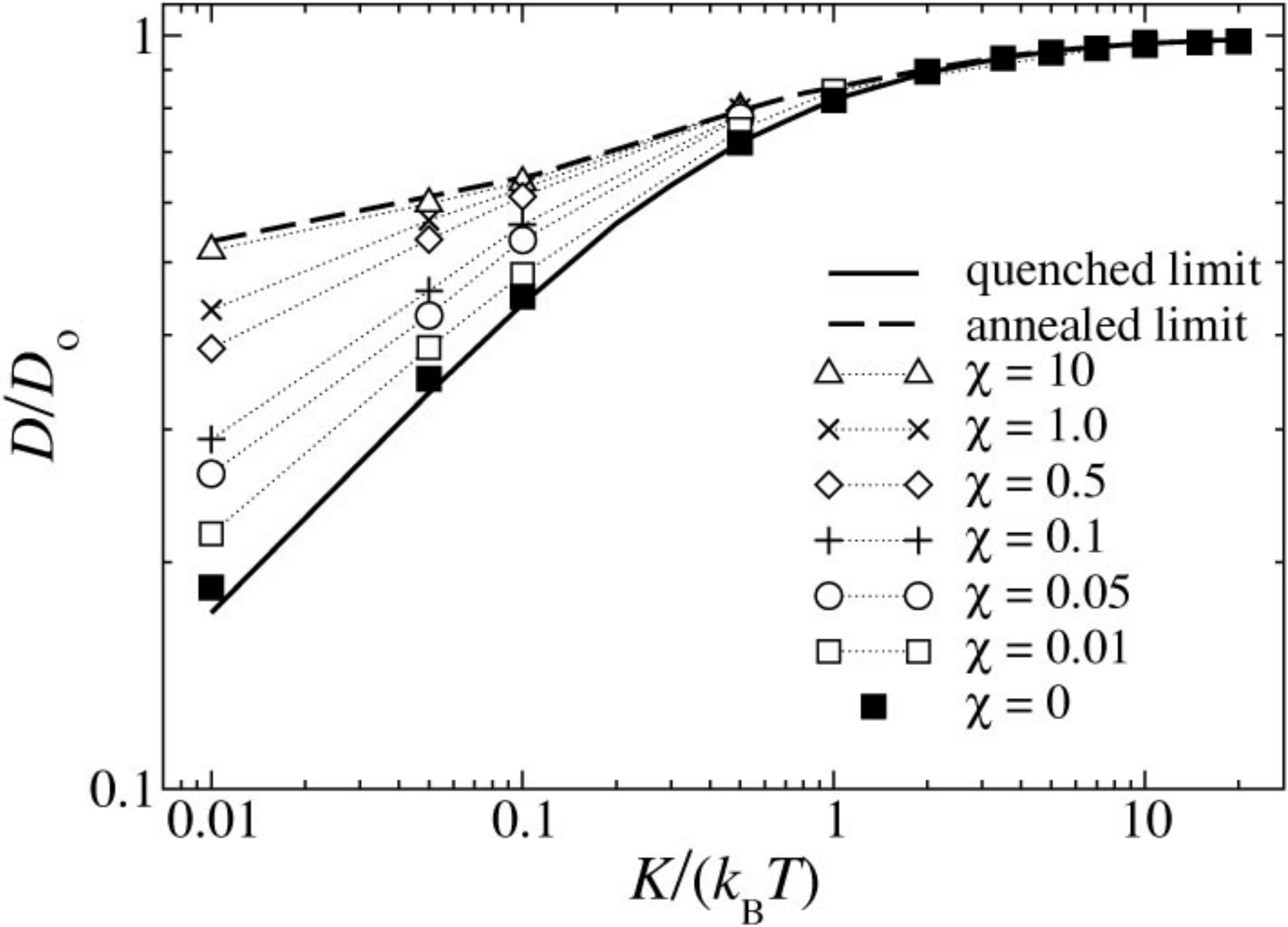}
\caption{Rescaled projected diffusion coefficient, $D/D_0$, of a freely diffusing particle on a dynamically
fluctuating elastic membrane as a function of the rescaled membrane bending rigidity 
$K/(k_{\mathrm{B}}T)$ for $M=32$ and surface tension $\sigma=0$. 
Symbols represent simulation data for various dynamical coupling parameters, $\chi$, spanning the whole
intermediate dynamical coupling regime from quenched membrane limit with $\chi=0$ (filled squared, same 
data as in Figure \ref{fig:DvsK}) to a rapidly fluctuating membrane with $\chi=10$ (open triangle-ups). 
The dashed curve is the annealed theoretical prediction (as calculated in Appendix \ref{app:annealed_theory}),  
while the solid curve is the quenched  area-scaling prediction, Eq. (\ref{eq:area_scaling}). Dotted lines are guide 
to eyes. Error-bars are removed for clarity (they are  typically about  the symbol size and comparable to or smaller than 
those shown in Fig. \ref{fig:DvsK}).  }
\label{fig:DvsK_varChi}
\end{center}
\end{figure}

\subsection{Simulation results}

In Figure \ref{fig:DvsK_varChi}, we show simulation results for the ratio of the projected 
diffusion coefficient to the curvilinear diffusion coefficient, $D/D_0$, as  a function of  the
membrane bending rigidity with a number of different values chosen for the dynamical 
coupling parameter, $\chi$. The case of $\chi=0$ (filled squares) represents a strictly quenched 
membrane, with results reproduced from Section  \ref{subsubsec:sim_quenched_memb}. 
As mentioned before, the quenched data agree with the area-scaling prediction (solid curve). 
(In the Figure, we use a log-log plot and explore bending rigidities 
down to very small (unphysical for lipid bilayers) values to clearly display the
predicted trends.  At large $K$, both quenched and annealed
results converge to within just a few percent.) 

It is clearly seen that, as the dynamical coupling parameter increases, the projected 
diffusion coefficient deviates from and becomes larger than 
its quenched value, indicating that dynamical fluctuations of the
surface enhance particle diffusion. 
This is to be expected based on very general considerations that imply 
 fast surface fluctuations help to excite the particle through potential barriers
 more easily as it moves over the surface \cite{Halle}. 
 For larger values of $\chi$ (stronger surface fluctuations), the data tend to the theoretical annealed 
 prediction (shown by a dashed curve in Figure \ref{fig:DvsK_varChi}) exhibiting a reasonably 
 good quantitative agreement for $\chi>10$ (see below).  
 The calculation of the annealed curve is presented in Appendix \ref{app:annealed_theory}. 
 It slightly differs from the approximate analytical results 
 previously obtained in Refs. \cite{Halle,Seifert} as we explain in the Appendix.

\begin{figure}[!t]
\begin{center}
\includegraphics[angle=0,width=8cm]{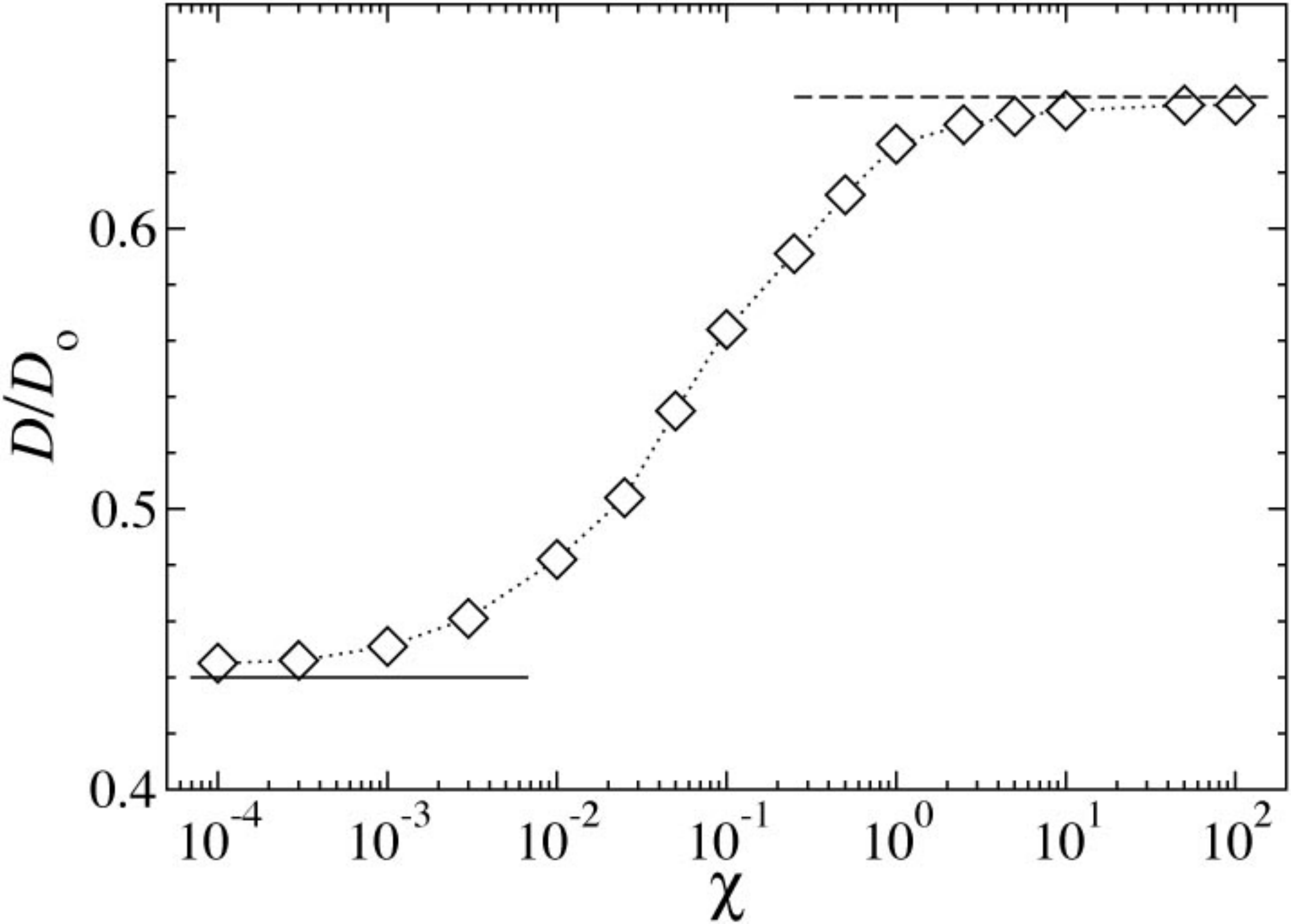}
\caption{Rescaled projected diffusion coefficient, $D/D_0$, of a freely diffusing particle on a dynamically
fluctuating elastic membrane as a function of the dynamical coupling parameter $\chi = k_{\mathrm{B}}T/(4\eta D_0 L)$
at fixed membrane bending rigidity of $K/(k_{\mathrm{B}}T)=0.1$ for $M=32$
and surface tension $\sigma=0$.
Symbols represent simulation data, the dashed horizontal line specifies the corresponding theoretical annealed value
(Appendix \ref{app:annealed_theory}), 
while the solid line represents the quenched area-scaling value (dotted curve is guide to eyes). 
A smooth crossover is observed between the asymptotic annealed and quenched  predictions. 
Error-bars are comparable to the symbol size.}
\label{fig:DvsChi}
\end{center}
\end{figure}

 In order to examine the crossover behavior from the quenched to annealed disorder limit, we 
 show in Figure \ref{fig:DvsChi} the simulated diffusion coefficients as 
 a function of the dynamical coupling parameter at a fixed value of the bending rigidity $K/(k_{\mathrm{B}}T)=0.1$. 
 Here we change $\chi$ over a range of six orders of magnitude spanning the transition from
quenched to annealed regimes.
The data exhibit a smooth crossover from the quenched area-scaling
value (shown by a solid horizontal line) to the predicted annealed value (shown by a dashed horizontal line). 
The agreement is almost perfect at both converging ends within the statistical error-bars of the data (comparable to
the symbol size, amounting to relative errors of about 2\%). 
Note that in these regions of $\chi$ (e.g., for $\chi>1$ or $\chi<10^{-3}$), the data exhibit a very weak dependence 
on $\chi$. As previously mentioned, this directly translates to a weak dependence of the ratio $D/D_0$ 
on the curvilinear diffusion coefficient, $D_0$, when all other parameters are fixed (see Eq. (\ref{eq:D_rescaledform})).  
In other words, in either limit $\chi \rightarrow 0$ or $\chi \rightarrow \infty$, $D/D_0$
depends only upon the dimensionless parameters, $K/(k_{\mathrm{B}}T)$ and $M$ 
(as well as $\sigma L^2/(k_{\mathrm{B}}T)$ when present).
The practical definition of these limits is implied by the curve in Figure \ref{fig:DvsChi}.

We find the same trend for all values of the bending rigidity as may be seen from Figure \ref{fig:DvsK_varChi}. 
Qualitatively, one can think of the $\chi$ value where $D/D_0$ reaches the mean value of the quenched and annealed
predictions as a conventionally chosen crossover point beyond or below which the system 
is dominated by either quenched or annealed features as described in Section \ref{subsubsec:time_scales}. 
In Figure \ref{fig:DvsChi}, this corresponds to a value of $\chi\sim 0.1$. For smaller
bending rigidities, this crossover point tends to larger values of $\chi$ (as may already be seen from
Figure \ref{fig:DvsK_varChi}), reflecting stronger quenched effects. 
A reverse trend is observed for larger bending rigidities. 

\subsection{Discussion of experimental systems}
The following numbers represent typical values for integral membrane proteins diffusing in
lipid bilayers \cite{Boal_Book} (but are specific to the case of diffusion of band 3 protein
on the surface of the human red blood cell in the absence of osmotic stress and neglecting
interactions with the cellular cytoskeleton \cite{Brown}):  
$D_0\simeq 0.5$~$\mu{\mathrm{m}}^2/{\mathrm{s}}$, $\eta = 0.06$ poise, $K/(k_{\mathrm{B}}T) \simeq 5$ and $\sigma=0$.  For length
scales around $2\pi q^{-1}= L\simeq 100$~nm,  these numbers translate to $\chi\simeq 3$ 
and the time scale ratio (\ref{eq:tau_ratio}) of $\tau^d_q/\tau^m_q\simeq 10^2$, which
is close to the annealed limit.  However, optical measurements are limited to (at best)
the diffraction limit of light, which is closer to $1~\mu \mathrm{m}$ and cells are typically on
the order of $10~\mu{\mathrm{m}}$ or larger.  For these larger membrane sizes of 
$L\simeq 1-10$~$\mu{\mathrm{m}}$, one finds $\chi \simeq 0.3-0.03$ or 
the time scale ratio of  $\tau^d_q/\tau^m_q\simeq 10-1$, which spans the 
intermediate regime.  While some experimental systems may
be adequately described by the annealed limit, other membrane systems, 
especially large ones, seem ill suited to such a description.  The general results
presented here will be of use in predicting $D$ for systems outside the annealed and/or quenched
regimes.



\section{Conclusion and discussion}
\label{sec:discussion}

We have analyzed the projected self-diffusion coefficient of a free Brownian particle on 
rough two-dimensional surfaces that may be static or fluctuating in time. 
This is achieved by establishing a position Langevin equation appropriate
for particle diffusion on weakly curved membranes represented in Monge gauge.  
The Langevin equation implicitly incorporates curvature and gradient effects that
hinder the diffusion of the particle once projected on a laboratory base plane. 
We have applied this equation numerically (using Brownian Dynamics simulation techniques) 
to several different examples of a static as well as a dynamic
surface. 

As shown, the dynamical state of the coupled particle-membrane system may be described
by a dynamical coupling parameter, $\chi =  k_{\mathrm{B}}T/(4\eta L D_0)$ (with $\eta$
being the medium viscosity, $L$ the system size, and $D_0$ the particle curvilinear diffusion coefficient), 
that controls the separation of characteristic time
scales for particle diffusion and membrane relaxation.  Qualitatively, it represents 
the ratio between the relaxation time of membrane undulations of typical energy scale $k_{\mathrm{B}}T$
and lateral range $L$
to the time spent by the particle to diffuse over such a length scale.  
This parameter specifies the dynamical nature of surface disorder experienced
by the diffusing particle with two well-defined limiting cases of quenched and 
annealed disorder being given by 
$\chi\rightarrow 0$ (static surface or fast diffusion) and $\chi\rightarrow \infty$ 
(fast-fluctuating surface or slow diffusion),  respectively.

The simulation methods presented in this work enable one to study the diffusion process
for the whole range of non-trivial intermediate couplings $\chi$, where the membrane and
particle diffusion time scales are comparable. 

In all quenched cases considered here, we find that the simulation data are best described
by the area-scaling prediction for the projected diffusion coefficient.  Indeed, to within our
statistical errors, the area-scaling result can be considered exact.   The area-scaling prediction has so far been 
considered only as a plausible scaling result \cite{Halle,Gov} and no rigorous proof is available, to our knowledge,
that demonstrates its validity in a systematic fashion. The variational bounds  and the 
effective medium approximation  discussed in detail in Ref. \cite{Halle} show notable deviations 
 from simulation data in all the studied cases. 
The present findings call for a more detailed analysis of diffusion over 
two-dimensional surfaces, and in particular, an analytical explanation that can clarify the unanticipated 
success of area-scaling in the quenched disorder limit.

In view of the fact that many biological structures exhibit large static undulations (as in the endoplasmic 
reticulum \cite{The-Cell} and the lateral cortex of auditory outer hair cells \cite{Boal_Book,OHCs}), 
it is important to realize that the annealed result may not be applied to these system.  Area scaling provides the appropriate  choice  for $D/D_0$. 

In the annealed limit ($\chi\gg 1$), we have shown that the simulation data for the projected diffusion 
coefficient agree well with analytical annealed limiting results as calculated in
Appendix \ref{app:annealed_theory}.  The annealed diffusion coefficients are higher than 
the corresponding quenched values. This is more pronounced for strongly roughened membranes that may 
be represented by small bending rigidity (and surface tension). 
For vanishing bending rigidity, the annealed diffusion coefficient tends to $D/D_0 = 1/2$, whereas the
corresponding quenched value tends to zero, reflecting a huge qualitative difference between the two in this limit. 
For large bending rigidities, the surface is almost planar and all different theoretical predictions converge to
$D/D_0 = 1$, with deviations of only a few percent for $K/(k_{\mathrm{B}}T)>1$. 
This is why  we have extensively studied the low bending rigidity regime in order to emphasize the physical mechanisms which affect the diffusion coefficient.

An important advantage of the present Langevin approach is that it can be easily generalized
to include particle-particle and membrane-particle interactions (as in the case of
active protein inclusions \cite{Lin-Gov2006}, where--unlike the present case--membrane dynamics
are directly influenced by forces exerted by the proteins). This will enable numerical investigation 
of  more complex and realistic models of biological membranes. 

As a final note, we comment on some of the physics absent from the present study.  This work considers
the influence of surface shape on the diffusion process solely through the geometric effect of
projection onto a base plane.  We have neglected any effects whereby surface shape may alter
the local character of the surface and the value of $D_0$.  For example, the curvature of a lipid
bilayer may influence the value of $D_0$ itself \cite{Gov} and we have neglected this effect.  In the
case of fluid lipid bilayers, it is clear that surface fluctuations must be accompanied by a local
flow of lipids as membrane material is moved about.  Such flows would be expected to drive
particle motion and lead to correlations between membrane fluctuations and protein transport.  
Our study, and the others we are aware of  \cite{Seifert,Gov}, completely neglect this effect.
The problem of coupled fluctuations and hydrodynamic flow within the bilayer is a seemingly
complex one, with only formal results available thus far \cite{lubensky_cai}.



\begin{acknowledgments}
We thank Hoda Boroudjerdi, David Dean, Nir Gov, Lawrence Lin and Phil Pincus for 
helpful discussions. This work was supported by the National 
Science Foundation (grant No. CHE-0321368 and grant No. CHE-0349196).  F.B. is an Alfred P. Sloan Research Fellow and a Camille Dreyfus Teacher-Scholar.

\end{acknowledgments}



\subsubsection*{Note Added in Proof}

An interesting paper by Reister-Gottfried et al. \cite{Reister2007} was published while this work was under review.
That paper also considers the problem of diffusion over an annealed flexible membrane.


\appendix

\section{Langevin equation in reduced representation} 
\label{app:rescaled}

In this Appendix we derive the rescaled functions that enter in the rescaled Langevin equation
for particle motion, Eq. (\ref{eq:Langevin_particle_res}).  As explained in the text, this follows
by simply plugging rescaled variables $\tilde x = x/L$, $\tilde t = D_0t/L^2$ and $\tilde h(\tilde x, \tilde y, \tilde t)=h(x, y, t)/L$ into the actual
Langevin equation (\ref{eq:Langevin}) (or simply multiplying both sides
of the equation by $L/D_0$). The  rescaled drift, $ \tilde v_i $, the rescaled multiplicative 
noise factor, $ \tilde \tau_{ij}$ (being the square-root of the inverse metric tensor, Eq. (\ref{eq:tau_ij})), 
may thus  be obtained  as 
\begin{equation}
  \tilde v_i\big[\tilde h(\{\tilde x_k\}, \tilde t)\big] = \frac{L}{D_0}\,  v_i, \,\,\,\,
  				\tilde \tau_{ij}\big[\tilde h(\{\tilde  x_k\}, \tilde t)\big] =  \tau_{ij}, 
\end{equation}
where we have explicitly indicated their dependence on space coordinates (particle position) 
through the surface height profile, $h(\{x_k\}, t) = h(x, y, t)$ (for $i, j, k=1,2$ in 2D). 
It is to be noted that the metric tensor $g_{ij}$, 
and therefore its inverse, $(\mathbf{g}^{-1})_{ij}$, and determinant, $g$, remain invariant under 
rescaling coordinates, e.g., $g_{ij}= \tilde g_{ij} \equiv \delta_{ij} + \tilde \partial_i \tilde h \, \tilde\partial_j \tilde  h$,
where $ \tilde \partial_i \equiv  \partial/\partial \tilde x_i$.  Hence using Eqs. (\ref{eq:drift_h}) and (\ref{eq:tau_ij}) we have
\begin{eqnarray}
  \tilde v_i& = &- \bigg[(\mathbf{g}^{-1})_{jk} \, \tilde \partial_j \tilde \partial_k \tilde h \bigg]\,\frac{ \tilde \partial_i \tilde h}{g}, \\
  \tilde \tau_{ij}&=&  \delta_{ij} - \frac{\tilde \partial_i \tilde h\, \tilde \partial_j \tilde  h}{g + \sqrt{g}}. 
\end{eqnarray} 

Likewise, the noise term, $\eta_i(t)$, in Eq. (\ref{eq:Langevin}) is rescaled as $\tilde \eta_i(\tilde t) = L \eta_i(t)/D_0$. 
Therefore, $\tilde \eta_i(\tilde t)$ is a Gaussian white noise with zero mean, $\langle \tilde \eta_i(\tilde t)\rangle=0$, and
variance $\langle \tilde \eta_i(\tilde t)\, \tilde \eta_j(\tilde t') \rangle = 2\delta_{ij}\,\delta(\tilde t -\tilde t')$. 

A similar scheme can be used to rescale the membrane dynamical equation (\ref{eq:q_memb_langevin_II}) as 
explained in the text. In this case, the rescaled noise term,  $\tilde \xi_{\tilde {\mathbf q}}(\tilde t)$, in 
Eq. (\ref{eq:q_memb_langevin_II_res}) is related to the actual noise term, $\xi_{{\mathbf q}}(t)$, in 
Eq. (\ref{eq:q_memb_langevin_II}), via $\tilde \xi_{\tilde {\mathbf q}}(\tilde t) = L\xi_{{\mathbf q}}(t)/\sqrt{D_0}$. 
Therefore, $\tilde \xi_{\tilde {\mathbf q}}(\tilde t)$ is a Gaussian white noise with 
$\langle \tilde \xi_{\tilde {\mathbf q}}(\tilde t)\rangle =0$ and variance   
$\langle\tilde \xi_{\tilde {\mathbf q}}(\tilde t) \, \tilde \xi_{\tilde {\mathbf q}'}(\tilde t') \rangle = 
2\delta_{\tilde {\mathbf q}, -\tilde {\mathbf q}'} \,\delta(\tilde t-\tilde t')$.

\section{Theoretical annealed prediction for the ratio $D/D_0$}
\label{app:annealed_theory}

As originally shown by Gustafsson et al.  \cite{Halle}, the diffusion coefficient ratio $D/D_0$ for a 
free particle diffusing on a fast-fluctuating surface is given by
\begin{equation}
  \frac{D}{D_0}\bigg|_{\mathrm{annealed}} = \frac{1}{2}\bigg(1+\bigg\langle \frac{1}{g}\bigg\rangle_h \bigg), 
  \label{eq:D_annealed}
\end{equation}
which has a form similar to the upper bound in the quenched limit, Eq. (\ref{eq:upper}), 
but here brackets denote an ensemble average over equilibrium configurations of the surface
as will be specified below. Gustafsson et al.  \cite{Halle} calculated this average for a generic {\em infinite} surface
 exhibiting an isotropic Gaussian distribution of surface gradients, $\nabla h$ (and
 thus a scalar diffusion coefficient).  For an elastic 
membrane of arbitrary size $L$ (and finite small-scale cut-off, $a$), 
the gradient distribution is in general anisotropic. In this Appendix, we demonstrate 
how the ensemble average in Eq. (\ref{eq:D_annealed}) can be calculated for this latter case.

The ensemble average of any function, $Q\big(\nabla h({\boldsymbol \rho})\big)$, of surface 
height gradient, $\nabla h$, at a given point ${\boldsymbol \rho} = (x, y)$, 
can be expressed as a functional integral over equilibrium membrane configurations as
\begin{equation}
 \big\langle Q\big(\nabla  h({\boldsymbol \rho})\big)\big\rangle_h = 
 		\int\! \frac{{\mathcal D}[h({\boldsymbol \rho}')]}{{\mathcal Z}_h} \, 
				Q\big(\nabla  h({\boldsymbol \rho})\big) \, 
					e^{-\beta {\mathcal H}[h({\boldsymbol \rho}')]}, 
	\label{eq:def0_averageQ}				
\end{equation}
where $\beta = 1/(k_{\mathrm{B}}T)$ and ${\mathcal Z}_h = \int\! {\mathcal D}[h]\exp[-{\mathcal H}/(k_{\mathrm{B}}T)]$
 is the membrane partition function. 
An example for $Q$ is the quantity $1/g$  appearing in Eq. (\ref{eq:D_annealed}), where $g \equiv 1+ (\nabla h)^2$. 
Denoting statistically probable values of $\nabla h$ by ${\mathbf u}$, we obtain a formally equivalent expression
for $\langle Q \rangle_h$ at a given point ${\boldsymbol \rho}$, i.e.
\begin{equation}
 \big\langle Q\big\rangle_h = \int\! {\mathrm{d}} {\mathbf u}\, \,
 			Q({\mathbf u}; {\boldsymbol \rho})\, P_u({\mathbf u}; {\boldsymbol \rho}),
\label{eq:def_averageQ}
 \end{equation}
where 
\begin{eqnarray}
 P_u({\mathbf u}; {\boldsymbol \rho})\! &=& \!
 		 \int\! \frac{{\mathcal D}[h({\boldsymbol \rho}')]}{{\mathcal Z}_h}  
				\, \delta\big({\mathbf u} - \nabla h({\boldsymbol \rho} )\big) \, 
					e^{-\beta{\mathcal H}[h({\boldsymbol \rho}')]} 
					\label{eq:P_u_integral}\\
 		\!&=& \! \bigg\langle \delta\big({\mathbf u} - \nabla h({\boldsymbol \rho} )\big) \bigg\rangle_h, 
\end{eqnarray}
is nothing but the probability distribution function of the height gradient, ${\mathbf u} = \nabla h$, at position ${\boldsymbol \rho}$. 
We thus first need to determine  $P_u({\mathbf u}; {\boldsymbol \rho})$ in order to calculate Eq. (\ref{eq:D_annealed}).  

We proceed by replacing the delta-function in Eq. (\ref{eq:P_u_integral}) by its Fourier representation, i.e.
\begin{equation}
\delta\big({\mathbf u} - \nabla h({\boldsymbol \rho} )\big) = 
		\int\! \frac{ {\mathrm{d}} {\boldsymbol \mu}}{(2\pi)^2}\, 
				\exp\!\bigg[ {\mathrm{i}} {\boldsymbol \mu}\cdot \big({\mathbf u} - \nabla h({\boldsymbol \rho} ) \big)\bigg], 
\end{equation}
where ${\boldsymbol \mu}$ is a Fourier conjugate of ${\mathbf u}$. Inserting this into Eq. (\ref{eq:P_u_integral}) and
using the explicit form of the Hamiltonian (\ref{eq:H}) (in Fourier representation), we reach at the following expression 
\begin{equation}
 P_u({\mathbf u}; {\boldsymbol \rho}) =  
 		 \int\! \frac{ {\mathrm{d}} {\boldsymbol \mu}}{(2\pi)^2} \, \,
		 		e^{ {\mathrm{i}} {\boldsymbol \mu}\cdot {\mathbf u} }\, \,
		 			{\mathcal I}\big({\boldsymbol \mu};{\boldsymbol \rho} \big), 
\label{eq:Pu_Iu}
\end{equation}
with ${\mathcal I}({\boldsymbol \mu};{\boldsymbol \rho})$ being the so-called generating function of 
height-gradient  moments, which is obtained as
\begin{equation}
  {\mathcal I}\big({\boldsymbol \mu};{\boldsymbol \rho} \big) = \int\! \frac{{\mathcal D}h_{\mathbf q}}{{\mathcal Z}_h}
  		\, e^{-\frac{1}{2 L^2}\sum_{{\mathbf q}} \big(\beta\Omega_{\mathbf q} |h_{\mathbf q}|^2  - 
			2 \,{\boldsymbol \mu}\cdot {\mathbf q} \, e^{ {\mathrm{i}}{\mathbf q}\cdot  {\boldsymbol \rho} } h_{\mathbf q}\big)}, 
\end{equation}
where ${\mathcal D}h_{\mathbf q} = \prod_{{\mathbf q}} {\mathrm{d}}h_{\mathbf q}$ and 
$\Omega_{\mathbf q} = Kq^4+\sigma q^2$. The above Gaussian integral can be evaluated and it follows that
the dependence on the position  of the generating function, ${\mathcal I}$, drops out yielding
\begin{equation}
   {\mathcal I}\big({\boldsymbol \mu};{\boldsymbol \rho} \big) =  {\mathcal I}\big({\boldsymbol \mu} \big) = 
   			\exp\!\bigg[\!-\frac{1}{2 \beta L^2}\sum_{{\mathbf q}} \frac{({\mathbf q}\cdot  {\boldsymbol \mu})^2}{\Omega_{\mathbf q}} \bigg]. 
\end{equation}
Putting this expression into Eq. (\ref{eq:Pu_Iu}), one finds after taking the Gaussian integral over ${\boldsymbol \mu}$ that
\begin{equation}
 P_u({\mathbf u}; {\boldsymbol \rho}) =  P_u({\mathbf u})  = 
 				\frac{ e^{ - \frac{1}{2}{\mathbf u} \cdot \hat {\mathbf A}^{-1}\cdot {\mathbf u} } }
							{2\pi \sqrt{ {\mathrm{det}}\hat {\mathbf A} } }, 
\label{Pu_anisotropic}
\end{equation}
where the elements of the $2\times2$ matrix $\hat {\mathbf A}$ read 
\begin{equation}
  (\hat {\mathbf A})_{ij} \equiv \big\langle  \partial_i h\, \partial_j h  \big\rangle =
  			 \frac{1}{\beta L^2} \sum_{{\mathbf q}} \frac{q_i q_j}{\Omega_{\mathbf q}}.
\end{equation}

In the limit of large  membrane size  $M=L/a\rightarrow \infty$, 
the cross terms (off-diagonal elements) of $\hat {\mathbf A}$ vanish  leading to isotropic 
height-gradient probability distribution used in Ref. \cite{Halle}, i.e.
\begin{equation}
   P_u({\mathbf u})  \rightarrow
 				\frac{1}{\pi\, \overline{u^2} }\, e^{ - {\mathbf u}^2/\overline{u^2}   }
\end{equation}
with the variance defined as $\overline{u^2}  = \langle(\nabla h)^2\rangle$. 

For a finite $M=L/a$, one can calculate desirable averages, Eq. (\ref{eq:def_averageQ}), using the 
anisotropic height-gradient distribution (\ref{Pu_anisotropic}). 
For the particular case of $\langle 1/g\rangle_h$ in 
Eq. (\ref{eq:D_annealed}), the computations need to be carried out numerically. 
The results for the predicted annealed ratio $D/D_0$, Eq. (\ref{eq:D_annealed}), 
at different bending rigidities (and zero surface tension) are 
shown in Figures \ref{fig:DvsK_varChi} and \ref{fig:DvsChi} (dashed lines). 
(Note that the annealed expression for $D$ in general involves additional
``off-diagonal" terms  containing cross-correlation functions, but they are negliggible for the 
membrane size chosen here--see Appendix \ref{app:AEFV} and Ref. \cite{Seifert}.)

Reister et al. \cite{Seifert} present a method to explicitly evaluate Eq. (\ref{eq:D_annealed}) for an elastic
membrane, but their scheme actually computes $\big\langle {\mathcal A}_\bot/\big(\int {\mathrm{d}}^2 x\, g\big)\big\rangle_h$
instead of $\langle 1/ g\rangle_h$, which in principle represents a different quantity. 
The results reported in Ref. \cite{Seifert} underestimate the annealed values of $D/D_0$ by up to 5-10\%
depending on the bending rigidity (with the larger errors being where the discrete summation over 
${\mathbf q}$-modes is approximated by an integral).

\section{Annealed limit: Adiabatic elimination of fast variables}
\label{app:AEFV}

The coupled membrane-particle dynamical equations (\ref{eq:Langevin_particle_res}) and (\ref{eq:q_memb_langevin_II_res})
as represented in rescaled units provide an appropriate analytical framework for a systematic perturbative analysis 
around the annealed-disorder limit. The time-scale separation in this limit is such that the membrane degrees of freedom 
represent the fast variables, whereas the  particle degrees of freedom constitute the slow variables in the system. 
Under this condition, one can use the methods described in literature to adiabatically eliminate
the fast variables and derive effective dynamical equations for the slow ones. 
 
Equations (\ref{eq:Langevin_particle_res}) and (\ref{eq:q_memb_langevin_II_res}) already indicate how this procedure should
be done in the present context by making use of $\chi$ as an expansion parameter. 
Following Risken  \cite{Risken}, we can develop a perturbative scheme much in the spirit of Born-Oppenheimer
approximation. The Fokker-Planck equation for the joint probability distribution of the set of variables 
$\{\tilde h_{\tilde {\mathbf q}}(\tilde t), \tilde x_i(\tilde t)\}$ may be written as 
\begin{equation}
  \dot{\tilde P}(\{\tilde x_i, \tilde h_{\tilde {\mathbf q}} \}, \tilde t) = 
  			\big( \hat O_x + \chi \hat O_h \big) \tilde P(\{\tilde x_i, \tilde h_{\tilde {\mathbf q}}\}, \tilde t), 
\label{eq:joint_FP}
\end{equation}
where the Fokker-Planck operators associated with $x$- and $h$-variables read
\begin{eqnarray}
   \hat O_x & =& -\frac{\partial}{\partial \tilde x_i}\, \tilde v_i + \frac{\partial^2}{\partial \tilde x_i \partial \tilde x_j} \, ({\mathbf g}^{-1})_{ij} \\
      \hat O_h & = & \sum_{\tilde {\mathbf q}} \bigg(  \tilde \omega_{\tilde {\mathbf q}}   \frac{\partial}{\partial  \tilde h_{\tilde{\mathbf q}} }
      			\tilde h_{\tilde {\mathbf q}}+ \tilde \Lambda_{\tilde {\mathbf q}}
      				\frac{\partial^2}{\partial \tilde h_{\tilde{\mathbf q}} \partial \tilde h_{-\tilde{\mathbf q}}  } \bigg). 
\end{eqnarray}

The idea here is to expand $\tilde P(\{\tilde x_i, \tilde h_{\tilde {\mathbf q}}\}, \tilde t)$ 
in terms of the eigen-functions of the  operator $\hat O_h$ (which are in the present context independent of $\{\tilde x_i\}$), i.e.
\begin{equation}
  \tilde P(\{\tilde x_i, \tilde h_{\tilde {\mathbf q}} \}, \tilde t) = \sum_{m=0}^\infty c_m(\{\tilde x_i\}, \tilde t) \, \phi_m(\{\tilde h_{\tilde {\mathbf q}}\}). 
 \label{eq:eigen_exp}
\end{equation}
Here we have assumed that a stationary solution as well as a discrete spectrum of eigen-functions, 
$\phi_n(\{\tilde h_{\tilde {\mathbf q}}\})$ (with $n\geq 0$), exist,  i.e.
\begin{equation}
   \hat O_h \, \phi_n(\{\tilde h_{\tilde {\mathbf q}}\}) = - \lambda_n\, \phi_n(\{\tilde h_{\tilde {\mathbf q}}\}),
\end{equation}
with $ \lambda_0=0$ being the stationary solution corresponding to the equilibrium distribution function of membrane,
$\phi_0(\{\tilde h_{\tilde {\mathbf q}} \}) = \prod_{\tilde {\mathbf q}}\tilde P_{\mathrm{eq}}(\tilde h_{\tilde {\mathbf q}})$.
Note that $\tilde P_{\mathrm{eq}}$ is given by Eq. (\ref{eq:P_memb_eq}). 
Inserting Eq. (\ref{eq:eigen_exp}) into the joint Fokker-Planck equation (\ref{eq:joint_FP}), one finds 
\begin{equation}
   \bigg(\frac{\partial}{\partial \tilde t} + \chi  \lambda_n\bigg)\, c_n  = \sum_{m=0}^\infty {\mathcal L}_{nm}\, c_m,
   \label{eq:cn}
\end{equation}
where ${\mathcal L}_{nm}(\{\tilde x_i\}) = \int\! {\mathcal D}\tilde h_{\tilde {\mathbf q}}\, \phi_n^\dagger\, \hat O_x \, \phi_m$ with $\phi_n^\dagger$
being the eigen-function of the adjoint operator $\hat O_h ^\dagger$, which is needed since the Fokker-Planck operator
is non-Hermitian \cite{Risken}. 

For large $\chi\gg1$, the time derivative may be neglected in Eq. (\ref{eq:cn}) for $n\geq 1$ in the long-time 
limit $\tilde t \gg (\chi \lambda_1)^{-1}$, and only kept in the equation governing $c_0$ (with $\lambda_0=0$). 
By doing so, one can easily derive a perturbative expression for the latter equation. 
But it turns out  that $c_0$ is nothing but the time-dependent probability distribution of the slow variable 
averaged over the fast-variable distribution  \cite{Risken}, i.e., 
$c_0 = \tilde {\mathcal P}(\{\tilde x_i\}, \tilde t) = \int\!  {\mathcal D}\tilde h_{\tilde {\mathbf q}}\, \tilde P(\{\tilde x_i, \tilde h_{\tilde {\mathbf q}} \}, \tilde t)$. 
Therefore, one ends up with the desired effective equation governing the probability distribution of particle position  
 on a fast-fluctuating membrane, which admits  the following perturbative form
\begin{equation}
  \dot{\tilde {\mathcal P}}(\{\tilde x_i\}, \tilde t) = \bigg( \hat O_x^{(0)} + \chi^{-1}\, \hat O_x^{(1)} +{\mathcal O}(\chi^{-2}) \bigg) \tilde {\mathcal P}(\{\tilde x_i\},  \tilde t), 
 \label{eq:Px_effective}
\end{equation}
with the zeroth-order term being given by 
\begin{equation}
   \hat O_x^{(0)} \equiv  {\mathcal L}_{00}(\{\tilde x_i\}) = -\frac{\partial}{\partial \tilde x_i}\, \langle\tilde v_i\rangle_h
   		 + \frac{\partial^2}{\partial \tilde x_i \partial \tilde x_j} \, \langle({\mathbf g}^{-1})_{ij}\rangle_h, 
\end{equation}
where the drift, $\langle\tilde v_i\rangle_h$, 
and the inverse metric tensor (rescaled diffusion matrix), 
$\langle({\mathbf g}^{-1})_{ij}\rangle_h$, 
are averaged over the equilibrium distribution function of membrane height
at  a given point as, e.g., defined in Eq. (\ref{eq:def0_averageQ}).  
The first-order operator in Eq. (\ref{eq:Px_effective}) reads
$\hat O_x^{(1)} = \sum_{n=1}^\infty {\mathcal L}_{0n}(\{\tilde x_i\})\, {\mathcal L}_{n0}(\{\tilde x_i\})/\lambda_n$. 

Note that the zeroth-order term in Eq. (\ref{eq:Px_effective}) becomes {\em exact} in the annealed limit  $\chi\rightarrow \infty$. 
Using explicit expressions given before for $\tilde v_i$ and $({\mathbf g}^{-1})_{ij}$
(and neglecting small off-diagonal terms for large $L$),
the annealed expression for $D/D_0$, Eq. (\ref{eq:D_annealed}), immediately follows.  
The effective Fokker-Planck equation for particle diffusion on an annealed surface was first derived in Ref. \cite{Seifert}
using a pre-averaging approximation, which  gives the effective 
Fokker-Planck equation (\ref{eq:Px_effective}) up to  the zeroth-order term. 
The preceding derivation places this result within a systematic framework and predicts
higher-order corrections.



\begin{thebibliography}{99}
\bibitem{single-value}
Since persistence lengths of typical biomembranes are much larger than cellular dimensions \cite{lipowsky,Boal_Book}, 
it is completely natural to assume a membrane surface without overhangs and hence a single-valued $h(x,y)$.
\bibitem{Note_co_contra} For the sake of presentation, we do not distinguish between co- and contra-variant 
variables in our notation.
\bibitem{Note_diagon} Note that ${\boldsymbol \tau}=({\mathbf g}^{-1})^{1/2}$ may be written in a symmetric fashion 
as  ${\boldsymbol \tau} = {\mathbf S} {\boldsymbol \Lambda}^{1/2} {\mathbf S}^T$, 
where ${\mathbf S}$ is the eigen-vector matrix of the positive-definite inverse metric tensor ${\mathbf g}^{-1}$, 
and ${\boldsymbol \Lambda}$ its diagonal eigen-value matrix, both of which can be calculated explicitly
in Monge gauge using Eq. (\ref{eq:g_inverse_monge}).
\end{thebibliography}
\end{document}